\documentclass[aps,prc,superscriptaddress,twoside,twocolumn,nofootinbib,showpacs]{revtex4}
\usepackage{amsmath}
\usepackage{amssymb}
\usepackage{graphicx}
\newcommand*{\sdot}[2]{#1\!\cdot\!#2}
\newcommand*{\fs}[1]{{#1\!\!\!/}}
\newcommand*{\hc}{\text{H.\,c.}}
\newcommand*{\ia}{{\text{int}}}
\newcommand*{\bare}{{\text{bare}}}
\newcommand*{\kinject}{\bullet}
\newcommand*{\ket}[1]{|#1\rangle}
\newcommand*{\bra}[1]{\langle#1|}
\begin{document}

\title{Gauge-invariant approach to meson photoproduction\\
 including the final-state interaction}
\author{H. Haberzettl}
 \affiliation{Center for Nuclear Studies, Department of Physics, The George
Washington University, Washington, DC 20052, USA}
 \affiliation{Institut f{\"u}r Kernphysik (Theorie), Forschungszentrum J{\"u}lich,
 D-52425 J{\"u}lich, Germany}
\author{K. Nakayama}
\affiliation{Department of Physics and Astronomy, University of Georgia,
Athens, GA 30602, USA} \affiliation{Institut f{\"u}r Kernphysik (Theorie),
Forschungszentrum J{\"u}lich, D-52425 J{\"u}lich, Germany}
\author{S. Krewald}
\affiliation{Institut f{\"u}r Kernphysik (Theorie), Forschungszentrum
J{\"u}lich, D-52425 J{\"u}lich, Germany}

\date{23 May 2006}

\begin{abstract}
A fully gauge-invariant (pseudoscalar) meson photoproduction amplitude off a
nucleon including the final-state interaction is derived. The approach based on
a comprehensive field-theoretical formalism developed earlier by one of the
authors replaces certain dynamical features of the full interaction current by
phenomenological auxiliary contact currents. A procedure is outlined that
allows for a systematic improvement of this approximation. The feasibility of
the approach is illustrated by applying it to both the neutral and charged pion
photoproductions.
\end{abstract}


\maketitle


\section{Introduction}

Ever since the pioneering work by Chew, Goldberger, Low, and Nambu on pion
production \cite{Chew}, the study of photo- and electroproduction of mesons off
nucleons has been utilized as one of the major research avenues to learn about
the excited states of the nucleon. In order to extract accurate information on
nucleon resonances, one needs --- in addition to precise and extensive
experimental data --- reliable reaction theories that allow one to disentangle
the resonance contributions from the background contributions to the
observables.

The extant descriptions of meson photoproduction reactions span a wide range of
different approaches (e.g., Chiral Perturbation Theory, tree-level effective
Lagrangians, $K$-matrix approach, etc.\
\cite{BKLM,Bernard:1991rt,Bernard:1992nc,Bernard,Feuster,Penner,Scholten,DHKT,Hanstein}).
The present work is based on the dynamical framework of meson-exchange models
of hadronic interactions \cite{Nozawa,Surya,Sato,Pascalutsa} in which the
composite nature of hadronic vertices is accounted for by so-called form
factors. Since, at present, our theoretical understanding of these vertex form
factors is rather incomplete, one usually parameterizes the vertex structure by
phenomenological functions (usually of monopole or dipole form) whose
parameters are adjusted to fit the data. The presence of such form factors
spoils gauge invariance of the photoproduction amplitude already when dressing
the bare tree level in this phenomenological manner. The inclusion of an
explicit hadronic final-state interaction (FSI) further complicates this
problem since the construction of the corresponding interaction current (where
the photon interacts with the hadronic structure \emph{within} the vertex)
consistent with the FSI requires the knowledge of the underlying dynamical
structure, a requirement that is impossible to be satisfied in an approach
where phenomenological form factors are employed. In order to maintain gauge
invariance in this situation, one needs to resort to finding prescriptions that
are consistent at a phenomenological level with the various dynamical models.
The existing prescriptions
\cite{Gross,Ohta,Antw,HH1,HH2,Kvinikhidze,DW,Borasoy:2005zg} do not, and indeed
cannot, provide a unique answer to this problem since manifestly transverse
currents --- that have no bearing on gauge invariance --- can always be added
to any given prescription. From a phenomenological point of view, therefore, it
is unavoidable that one seeks a prescription that works best in reproducing the
data (see, e.g., discussion in \cite{Usov}). A number of the existing
gauge-invariance preserving prescriptions have already been applied in this
manner in pion photoproduction \cite{Surya,Sato,Pascalutsa} as well as in
electroproduction \cite{Nozawa1,Caia} reactions.

However, most of the existing calculations based on phenomenological dynamical
models are actually not gauge invariant. In fact, they revert to a variety of
{\it ad hoc} recipes for the sole purpose of enforcing current conservation,
\begin{equation}
k_\mu M^\mu = 0~,
 \label{cc}
\end{equation}
when the production current $M^\mu$ is on-shell, but not the gauge-invariance
condition expressed by the generalized Ward--Takahashi (WT) identity
\cite{kazes,Antw,HH1}
\begin{align}
k_\mu M^\mu & =  - \ket{F_s\tau}S_{p+k}Q_iS^{-1}_p
                 + S^{-1}_{p'}Q_f S_{p'-k}\ket{F_u\tau}
          \nonumber \\
   &\qquad\mbox{}
          +  \Delta^{-1}_{p-p'+k}Q_\pi\Delta_{p-p'}\ket{F_t\tau}~,
 \label{gi}
\end{align}
which is an \emph{off-shell} condition. [This equation is repeated as
Eq.~(\ref{gi1}) below, where its details are explained.]

One of the few exceptions to this situation is the recent work by Pascalutsa
and Tjon \cite{Pascalutsa} where a fully gauge invariant pion photoproduction
amplitude has been constructed based on the Gross--Riska prescription
\cite{Gross}. This approach has been extended and applied as well to pion
electroproduction \cite{Caia}. The prescription of Ref.~\cite{Gross} also has
been applied to the nucleon-nucleon bremsstrahlung reaction \cite{Fred}. The
Gross--Riska procedure relies on the observation that vertices and propagators
always enter in the combination (vertex\,$\times$\,propagator) in a given
reaction amplitude. Therefore, a vertex form factor that depends only on the
momentum of the propagating (off-shell) particle can be incorporated into the
corresponding propagator instead of being associated with the vertex. Of
course, if more than one leg of the vertex belongs to an off-shell particle,
this restricts the phenomenological form factors to being separable functions
of the respective leg momenta. In addition, the form factors should be such
that they do not lead to unphysical behavior of the resulting propagators.
Gauge invariance is then fulfilled by constructing electromagnetic currents
that obey the WT identities with the respective hadronic propagators modified
by the inclusion of the form factors as described.\footnote{Note that the
electromagnetic vertices constructed in this way in Refs.\cite{Pascalutsa,Fred}
differ from each other by a transverse piece.}  At the tree level, this
prescription completely removes the form factors from the longitudinal part of
the reaction amplitude; i.e., only manifestly transverse parts of the
production current carry any form factor dependence.

In the present work, we construct a photoproduction amplitude based on the
field-theoretical approach given by Haberzettl~\cite{HH1}. The full formalism
is gauge-invariant as a matter of course. However, in view of its complexity
and high nonlinearity, its practical implementations require that some reaction
mechanisms need to be truncated and/or replaced by phenomenological
approximations. Our objective here is to preserve full gauge invariance, in the
sense of Eq.~(\ref{gi}), for this approximate treatment, but allowing for the
presence of explicit hadronic FSIs. This problem has been treated already in
Ref.~\cite{HH3} as a two-step procedure where the gauge-invariant treatment of
explicit FSIs was added on to an already gauge-invariant tree-level amplitude
that had been constructed according to the prescriptions given in
Ref.~\cite{HH2}. The present approach instead starts from the full amplitude
and derives a single condition for the mechanisms to be approximated that
follows directly from the generalized WT identity (\ref{gi}). It thus is more
general and not tied to any particular tree-level treatment. Moreover, we
present a general scheme that allows a systematic way of including more complex
reaction mechanisms into the procedure. At the lowest order, it is found that
the essential aspects of the results found in Ref.~\cite{HH3} remain true. In
contrast to the prescription of Ref.~\cite{Gross}, the present approach does
not impose any restriction on the type of the hadronic form factors that can be
used. Furthermore, the longitudinal part of the resulting reaction amplitude
retains these form factors even at the tree-level.

The present paper is organized as follows. In Sec.~II we present our approach
to construct a fully gauge invariant photoproduction amplitude. In Sec.~III we
illustrate the approach developed in sect.~II by applying it to the pion
photoproduction. Section~IV contains a summary with our conclusions. Some
details of the present approach as well as of its model application are given
in the appendices.


\begin{figure*}[t!]
\includegraphics[width=\textwidth,clip=]{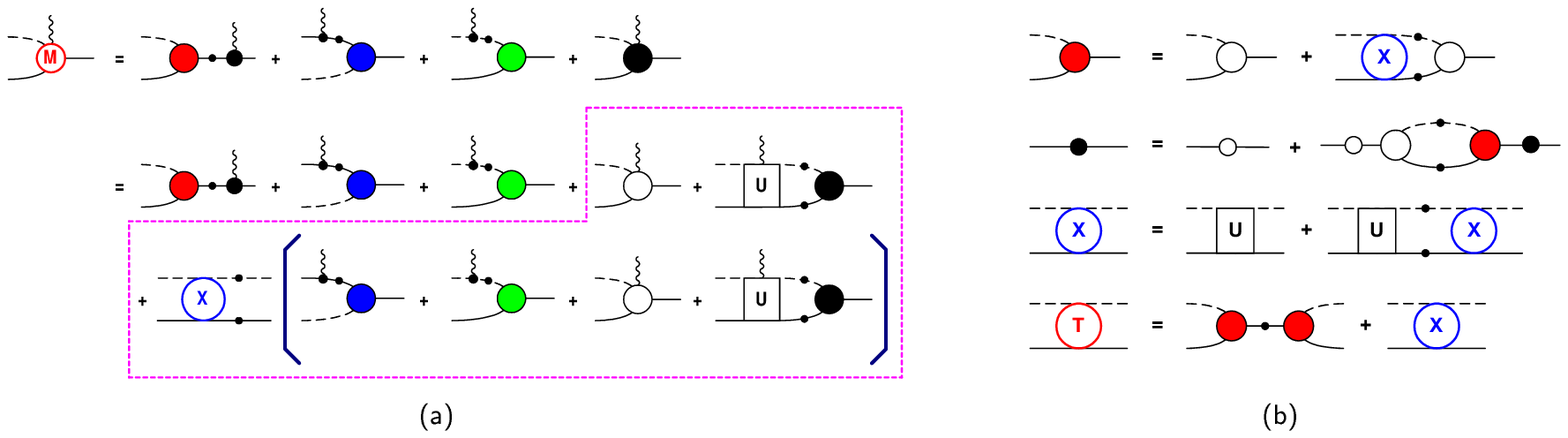}
  \caption{\label{diagrams1}  (Color online) Diagrammatic summary of the
  field-theory formalism of Ref.~\cite{HH1}. Time proceeds from right to left.
  ~~(a)~ Meson production current $M^\mu$. The first line corresponds to
  Eq.~(\ref{eq:Mmu_suti}) summing up, in that order, the $s$-, $u$-, and
  $t$-channel diagrams and the interaction current $M^\mu_{\rm int}$. (The
  different colors of the hadronic three-point vertices identify the $\pi NN$
  vertex in different kinematical situations.) The dynamical content of
  $M^\mu_{\rm int}$ is explicitly shown by the diagrams enclosed in the dashed
  box of the last two lines. This also includes, in the bottom line, the
  final-state interaction mediated by the nonpolar $\pi N$ amplitude $X$ that
  satisfies the integral equation shown in (b). The diagram element labeled $U$
  subsumes all exchange currents $U^\mu$ contributing to the process (see
  Fig.~\ref{fig:ucurr}). The diagram with open circle depicts the bare current
  $m^\mu_{\rm bare}$ (i.e., the Kroll--Ruderman term). ~~(b)~ Pion-nucleon
  scattering with dressed hadrons. The full $\pi N$-amplitude is denoted by
  $T$, with $X$ subsuming all of its nonpolar (i.e., non-$s$-channel)
  contributions. The latter satisfies the integral equation $X=U+UG_0X$
  depicted in the third line here, where the driving term $U$ sums up all
  nonpolar irreducible contributions to $\pi N$-scattering, i.e., all
  irreducible  contributions which do not contain an $s$-channel pole (see
  Ref.~\cite{HH1} for full details). --- In both parts (a) and (b), diagram
  elements with open, unlabeled circles describe bare quantities, and solid
  circles (or circles filled with colors) denote the corresponding dressed
  vertices and propagators.}
\end{figure*}

\section{Formalism}

In the following, for definiteness, we will explicitly consider the production
of pions off the nucleon, i.e., $\gamma+N\to \pi+N$, but the formalism will of
course apply equally well to the photoproduction (or electroproduction) of any
pseudoscalar meson. Moreover, at intermediate stages of the reaction, we will
ignore other mesons or baryonic states since they are irrelevant for the
problem at hand, namely how to preserve gauge invariance in the presence of
FSI.

\begin{figure}[t!]\centering
  \includegraphics[width=.9\columnwidth,clip=]{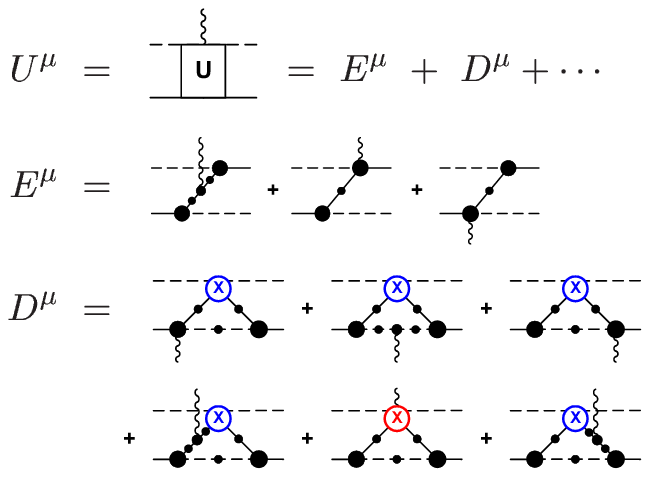}
  \caption{\label{fig:ucurr}%
  (Color online) Exchange-current contributions subsumed in $U^\mu$ grouped by
  the topological properties of the underlying $\pi N$ irreducible hadron
  contributions: The diagrams subsumed in $E^\mu$ are based on the photon
  attaching itself \emph{internally} in all possible ways to a simple hadron exchange
  graph, whereas $D^\mu$ subsumes the analogous contributions arising from a
  triangular hadron graph; more complex structures are not shown explicitly.
   Note that the second and third diagrams of $E^\mu$, and the first and third diagrams
   of $D^\mu$ explicitly contain the full \emph{off-shell} interaction current
  $M^\mu_\ia$. Implicitly it is contained in many more places. (The middle
  diagram in the second row  of $D^\mu$ contains the diagram $X^\mu$ where the
  photon is attached to the \emph{internal} structure of the non-polar $\pi N$
  amplitude $X$. The details of this mechanism are irrelevant for the present
  considerations; they can be found in   Ref.~\cite{HH1}.)}\vspace{-3mm}
\end{figure}

\begin{figure}[t!]\centering
  \includegraphics[width=.9\columnwidth,clip=]{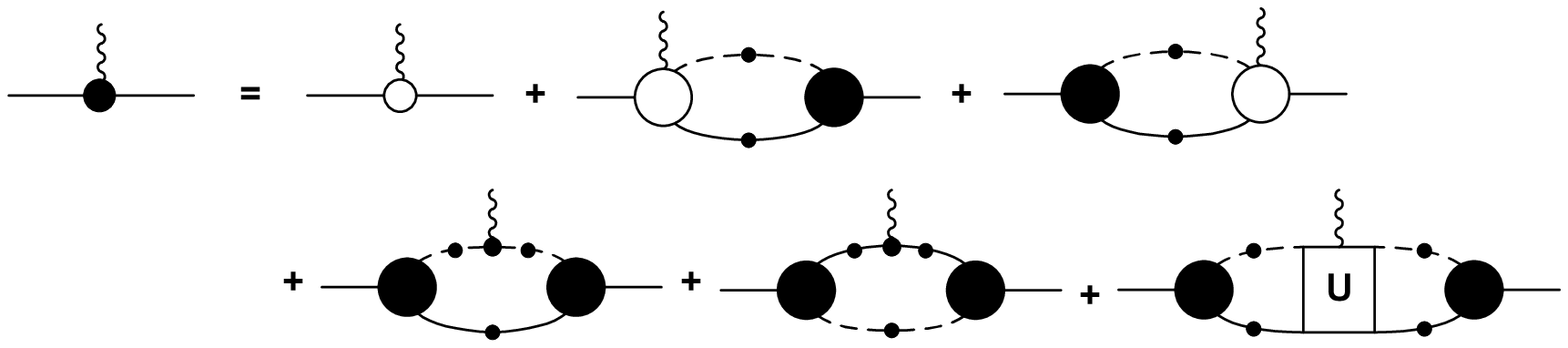}
  \caption{\label{fig:Ncurrent}%
  Diagrammatic representation of the nucleon's electromagnetic current
  obtained by attaching a photon to the second line of diagrams in
  Fig.~\ref{diagrams1}(b). The terms correspond to those of Eq.~(\ref{dvrtx}).}
\end{figure}

As mentioned, our approach is based on the field-theory formalism of
Ref.~\cite{HH1}. For the present purpose, however, we do not need to
recapitulate its full details. Instead, we employ the summarizing diagrams of
Figs.~\ref{diagrams1}, \ref{fig:ucurr}, and \ref{fig:Ncurrent}. As is
well-known \cite{Chew}, the photoproduction current $M^\mu$ can be broken down
according to
\begin{equation}
M^\mu = M^\mu_s+M^\mu_u+M^\mu_t+M^\mu_\ia~,
 \label{eq:Mmu_suti}
\end{equation}
where the first three terms describe the coupling of the photon to external
legs of the underlying $\pi NN$ vertex (with subscripts $s$, $u$, and $t$
referring to the appropriate Mandelstam variables of their respective
intermediate hadrons). These terms are relatively straightforward and easy to
implement in a practical application. However, the last term, the interaction
current $M^\mu_\ia$, where the photon couples inside the vertex, explicitly
contains the hadronic FSI; its structure is, therefore, more complex than that
of any of the first three terms. We read off the diagrams enclosed by the
dashed box of Fig.~1(a) that
\begin{align}
M^\mu_\ia &= m^\mu_\bare +U^\mu
G_0\ket{F\tau}\nonumber\\
&\quad\mbox{} +XG_0\Big(M^\mu_u+M^\mu_t+m^\mu_\bare +U^\mu
G_0\ket{F\tau}\Big)~,
\end{align}
where $m^\mu_\bare$ is the bare Kroll--Ruderman contact current, $U^\mu$
subsumes all possible exchange currents (see Fig.~\ref{fig:ucurr}), $G_0$
describes the intermediate $\pi N$ two-particle propagation, and the FSI is
mediated by the nonpolar part $X$ of the $\pi N$ $T$ matrix. Following
Ref.~\cite{HH1}, the notation $\ket{F\tau}$ is used for the dressed $N\to\pi N$
vertex (including its full coupling-operator structure); for $N\pi \to N$ with
the pion leg reversed, we use $\bra{F\tau}$. The isospin operator $\tau$ (with
its component index suppressed) is pulled out of the vertex explicitly for
later convenience [see Eq.~(\ref{eq:chargedef})]. The vertex obeys the equation
\begin{equation}
\ket{F\tau} = \left(1 + X G_0\right) \ket{F_\bare \tau} \ , \label{dvrtx_NNpi}
\end{equation}
where $\ket{F_\bare \tau}$ denotes the bare $\pi NN$ vertex.

Note that the $NN\gamma$ electromagnetic vertices appearing in
Fig.~\ref{diagrams1}(a), are also fully dressed vertices given by
\begin{align}
\Gamma^\mu_N & = {\Gamma^\mu_{N}}_\bare + \bar{m}^\mu_\bare G_0\ket{F\tau}\nonumber\\
&\quad\mbox{} +\bra{F\tau} G_0\Big(m^\mu_\bare +M^\mu_t+M^\mu_u+U^\mu
G_0\ket{F\tau}\Big)~, \label{dvrtx}
\end{align}
and illustrated diagrammatically in Fig.~\ref{fig:Ncurrent}.
${\Gamma^\mu_{N}}_\bare$ and  $\bar{m}^\mu_\bare$ denote the bare vertices for
$\gamma N\to N$ and $\gamma\pi N \to N$ (i.e., $\bar{m}^\mu_\bare$ is the
Kroll--Ruderman current $m^\mu_\bare$ with the pion leg reversed). Also, the
nucleon propagator, $S$, illustrated in the second row from the top in
Fig.~\ref{diagrams1}(b), is fully dressed according to
\begin{equation}
S^{-1} = S^{-1}_\bare -\bra{F_\bare \tau} G_0 \ket{F\tau} \ ,
\end{equation}
such that the WT identity as expressed by Eq.~(\ref{WTN}) below is satisfied.
$S_\bare$ stands for the bare nucleon propagator.

As alluded to in the Introduction, the production current $M^\mu$ is gauge
invariant if its four-divergence satisfies the generalized WT identity
\cite{HH1,kazes,Antw}
\begin{align}
k_\mu M^\mu & =  - \ket{F_s\tau}S_{p+k}Q_iS^{-1}_p
                 + S^{-1}_{p'}Q_f S_{p'-k}\ket{F_u\tau}
          \nonumber \\
   &\qquad\mbox{}
          +  \Delta^{-1}_{p-p'+k}Q_\pi\Delta_{p-p'}\ket{F_t\tau}~,
 \label{gi1}
\end{align}
where $p$ and $k$ are the four-momenta of the incoming nucleon and photon,
respectively, and $p'$ and $q$ are the four-momenta of the outgoing nucleon and
pion, respectively, related by momentum conservation $p'+q=p+k$. $S$ and
$\Delta$ are the propagators of the nucleons and pions, respectively, with
their subscripts denoting the available four-momentum for the corresponding
hadron; $Q_i$, $Q_f$, and $Q_\pi$ are the initial and final nucleon and the
pion charge operators, respectively. The index $x$ at $[F_x\tau]$ labels the
appropriate kinematic situation for $\pi NN$ vertex in the $s$-, $u$-, or
$t$-channel diagrams of Fig.~\ref{diagrams1}. This is an \emph{off-shell}
condition. In view of the inverse propagators appearing in each term here, if
all external hadronic legs are on-shell, this reduces to
\begin{equation}
k_\mu M^\mu =0
 \qquad \text{(hadrons on-shell)}~,
 \label{cc1}
\end{equation}
which describes current conservation.

Physically relevant, of course, is only current conservation. However, the
reason one must satisfy the off-shell condition (\ref{gi1}) for gauge
invariance to hold true is the requirement to have consistency across all
elements of the underlying reaction dynamics. In view of the inherent
nonlinearity of the process (due to the fact that the number of pions is not
conserved), the elements contributing to the full amplitude $M^\mu$ couple back
into themselves nonlinearly \cite{HH1}: For example, as can be seen from
Fig.~\ref{fig:ucurr}, the sum of exchange currents $U^\mu$ internally also
contains the interaction current $M^\mu_\ia$ in several places, with at least
one hadron leg off-shell \emph{even if all external hadrons are taken
on-shell}. It is then found that it is \emph{not} possible to achieve current
conservation consistently unless the current satisfies the off-shell condition
(\ref{gi1}), which translates into the condition (\ref{eq:GCMint}) for the
interaction current given below.

The electromagnetic currents for the nucleons and the pions, $\Gamma^\mu_N$ and
$\Gamma^\mu_\pi$, respectively, satisfy the WT identities
\begin{subequations}\label{eq:WTpiN}
\begin{align}
k_\mu \Gamma_N^\mu(p',p) &=S^{-1}_{p'}Q_N-Q_NS^{-1}_p~,
 \label{WTN}
\\
k_\mu \Gamma_\pi^\mu(q',q) &=\Delta^{-1}_{q'}Q_\pi-Q_\pi\Delta^{-1}_q~,
 \label{WTpi}
\end{align}
\end{subequations}
where the four-momentum relations $p'=p+k$ and $q'=q+k$ hold. It is therefore
possible to replace the generalized WT identity (\ref{gi1}) by the
\emph{equivalent} gauge-invariance condition
\begin{align}
k_\mu M^\mu_\ia & = - \ket{F_s\tau} Q_i +Q_f \ket{F_u\tau} e_f +Q_\pi \ket{F_t
\tau}
 \nonumber\\[1ex]
 &\equiv - \ket{F_s} e_i +\ket{F_u} e_f
   +\ket{F_t} e_\pi~,
 \label{eq:GCMint}
\end{align}
where the operators
\begin{equation}
e_i = \tau Q_i~,
 \quad
e_f = Q_f \tau ~,
 \quad\text{and}\quad
e_\pi = Q_\pi \tau
 \label{eq:chargedef}
\end{equation}
describe the respective hadronic charges in an appropriate isospin basis
(component indices and summations are suppressed here), i.e, apart from some
numerical factors, the $e_x$ are essentially given by the charges of the
respective particles. Charge conservation for the production process then
simply reads
\begin{equation}
e_i=e_f+e_\pi~.
\end{equation}
In the following, it is more convenient to use the condition (\ref{eq:GCMint}),
instead of (\ref{gi1}), together with (\ref{eq:WTpiN}).

We emphasize here that if the single-particle electromagnetic currents satisfy
the WT identities (\ref{eq:WTpiN}) and if the four-divergence of the
interaction current is given by (\ref{eq:GCMint}), then the corresponding
production amplitude $M^\mu$ will be gauge-invariant as a matter of course
\emph{even if propagators and vertices have been subjected to approximations}.

\subsection{Preserving gauge invariance}

\begin{figure}[t!]
  \includegraphics[width=.5\columnwidth,clip=]{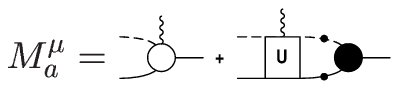}
  \caption{\label{fig:MAcurrent}%
  Diagrammatic representation of Eq.~(\ref{eq:Madef}).}
\end{figure}

The preceding considerations are completely general. In practical applications,
however, one will not be able to calculate all mechanisms that contribute to
the full reaction dynamics and one must make some approximations. This is
particularly true for the complex mechanisms that enter $U^\mu$, as depicted in
Fig.~\ref{fig:ucurr}. Approximations should preserve the gauge invariance of
the amplitude. To see how this can be done, let us define
\begin{equation}
M^\mu_a = m^\mu_\bare +U^\mu G_0 \ket{F\tau}~,
 \label{eq:Madef}
\end{equation}
which is shown in Fig.~\ref{fig:MAcurrent}, and  thus write
\begin{equation}
M^\mu_\ia = M^\mu_a +XG_0\Big(M^\mu_u+M^\mu_t+M^\mu_a\Big)~,
 \label{eq:Mint_Ma}
\end{equation}
and recast the gauge-invariance condition (\ref{eq:GCMint}) as a condition for
$M^\mu_a$. This immediately produces
\begin{align}
k_\mu M^\mu_a &= (1-UG_0)\big[- \ket{F_s} e_i +\ket{F_u} e_f
   +\ket{F_t} e_\pi\big]
 \nonumber\\[1ex]
 &\mbox{}\qquad
 -UG_0 \big[k_\mu (M^\mu_u +M^\mu_t)\big]~,
 \label{eq:GIMa1}
\end{align}
where
\begin{equation}
(1+XG_0)^{-1} = 1-UG_0
\end{equation}
and
\begin{equation}
U=\left(1+XG_0\right)^{-1}X
\end{equation}
were used, with $U$ being the sum of all non-polar hadronic driving terms [cf.\
third line of diagrams in Fig.~\ref{diagrams1}(b)]. In the last term of
(\ref{eq:GIMa1}) obviously only the \emph{non-transverse} parts of the $u$- and
$t$-channel currents $M^\mu_u$ and $M^\mu_t$ will contribute. Denoting those
respectively by $m^\mu_u$ and $m^\mu_t$, i.e.,
\begin{equation}
k_\mu(M^\mu_u-m^\mu_u)=0 \quad\text{and}\quad k_\mu(M^\mu_t-m^\mu_t)=0~,
\end{equation}
we finally have
\begin{align}
k_\mu M^\mu_a &= (1-UG_0)\big[- \ket{F_s} e_i +\ket{F_u} e_f
   +\ket{F_t} e_\pi\big]
 \nonumber\\[1ex]
 &\mbox{}\qquad
 -k_\mu UG_0 (m^\mu_u +m^\mu_t)
 \label{eq:GIMa2}
\end{align}
as the necessary condition that $M_a^\mu$ must satisfy so that $M^\mu_\ia$
yields the gauge-invariance condition (\ref{eq:GCMint}). Equation
(\ref{eq:GIMa2}) is exact --- no approximation has been made up to here.

\subsubsection{Approximating $M^\mu_a$}\label{sec:Mmua}

The structure of the preceding condition suggests the following approximation
strategy. The condition evidently is satisfied if we now approximate $M^\mu_a$
by
\begin{equation}
M_a^\mu = (1-UG_0) M^\mu_c - UG_0 (m^\mu_u +m^\mu_t) +T^\mu~,
 \label{eq:Ma_approx}
\end{equation}
where $M^\mu_c$ can be \emph{any} contact current satisfying
\begin{equation}
k_\mu M^\mu_c = - \ket{F_s} e_i +\ket{F_u} e_f
   +\ket{F_t} e_\pi
 \label{eq:McGI}
\end{equation}
and $T^\mu$ is an undetermined transverse \emph{contact} current that is
unconstrained by the four-divergence (\ref{eq:GIMa2}). With the choice
(\ref{eq:Ma_approx}), the corresponding approximate $M^\mu_\ia$ is then easily
found from (\ref{eq:Mint_Ma}) as
\begin{align}
M^\mu_\ia &=M_c^\mu +T^\mu
 \nonumber\\[1ex]
 &\mbox{}\quad+
 XG_0\big[(M^\mu_u-m^\mu_u)+(M^\mu_t-m^\mu_t)+T^\mu\big]~.
 \label{eq:MiaMc}
\end{align}
In this scheme, therefore, the choice one makes for $M^\mu_c$ (and $T^\mu$)
corresponds to an implicit approximation of the full dynamics contained in the
right-hand side of Eq.~(\ref{eq:Madef}). Moreover, beyond this actual choice,
the only explicit effect of the FSI $X$ is from explicitly \emph{transverse}
loop contributions, which is precisely the same result that was found in
Ref.~\cite{HH3}. Thus it follows that
\begin{equation}
k_\mu M^\mu_\ia= k_\mu M^\mu_c
\end{equation}
and this approximate interaction current then obviously satisfies the
gauge-invariance condition (\ref{eq:GCMint}).

Equations  (\ref{eq:Ma_approx}) and (\ref{eq:MiaMc}), together with the
prescriptions for $M^\mu_c$ and $T^\mu$ as given in the following two
subsections, are the main results of the present work.

Note that the choice of $T^\mu$, while it has no bearing on the gauge
invariance itself, will have a direct effect on how, if at all, the FSI enters
the approximate treatment. For example, putting for the moment
\begin{equation}
T^\mu = -UG_0\big[(M^\mu_u-m^\mu_u)+(M^\mu_t-m^\mu_t)\big]
 \label{eq:Tchoice1}
\end{equation}
simply provides
\begin{equation}
M^\mu_\ia=  M^\mu_c~.
\end{equation}
Therefore, this particular choice completely eliminates the explicit occurrence
of the FSI and, for phenomenological choices of $M^\mu_c$, such as
Eq.~(\ref{eq:MCbeta}) below, this corresponds to the tree-level approximation
where the full interaction current is replaced by a phenomenological contact
current. We emphasize, however, that we do not advocate actually using
Eq.~(\ref{eq:Tchoice1}). This particular (extreme) choice merely illustrates
that the undetermined $T^\mu$ may contain pieces that may be capable of
partially undoing the explicit inclusion of the FSI in (\ref{eq:MiaMc}). In
Sec.~\ref{sec:transcontact} we introduce a phenomenological procedure for
obtaining $T^\mu$ from the data.

With the approximation (\ref{eq:Ma_approx}) for $M^\mu_a$, the dressed
$NN\gamma$ vertex (\ref{dvrtx}) becomes
\begin{align}
\Gamma^\mu_N & = \Gamma^\mu_{N(\bare)} + \bar{m}^\mu_\bare G_0\ket{F\tau}\nonumber\\
&\quad\mbox{} +\bra{F_\bare \tau} G_0\Big(M^\mu_c+m^\mu_u+m^\mu_t\Big) \nonumber\\
&\quad\mbox{} +\bra{F\tau} G_0\Big[\left(M^\mu_u-m^\mu_u\right) +
\left(M^\mu_t-m^\mu_t\right)+T^\mu\Big]~, \label{dvrtx_approx}
\end{align}
where Eq.~(\ref{dvrtx_NNpi}) has been used. To the extend that the approximated
$M^\mu_a$ (\ref{eq:Ma_approx}) fulfills the same condition (\ref{eq:GIMa2})
satisfied by the exact current $M^\mu_a$, the above $NN\gamma$ vertex also
satisfies the same WT identity (\ref{WTN}) obeyed by the exact dressed vertex
given by Eq.~(\ref{dvrtx}).

\subsubsection{Choosing $M^\mu_c$}

The phenomenological choice that we make here for $M^\mu_c$ is a variant of the
procedure proposed in Refs.~\cite{HH2,HH1} that is more general than what was
suggested in \cite{HH3}. We parameterize the $\pi NN$ vertices by
\begin{equation}
F_x = g_\pi \gamma_5 \left[\lambda+(1-\lambda)
\frac{\fs{q}_\pi}{m+m'}\right]f_x~,
 \label{eq:Fdef}
\end{equation}
where $x=s$, $u$, or $t$ indicates the kinematic context, $g_\pi$ is the
physical coupling constant, $m$ and $m'$ are the nucleon masses before and
after the pion is emitted/absorbed and the parameter $\lambda$ allows for the
mixing of pseudoscalar (PS: $\lambda=1$) and pseudovector (PV: $\lambda=0$)
contributions. For simplicity, the functional dependence $f_x$ of the vertex
(which depends on the squared four-momenta of all three legs) is chosen as
common to both PS and PV, and it is normalized to unity if all vertex legs are
on-shell. We define then an auxiliary current
\begin{align}
  C^\mu
 &=  -e_\pi\frac{(2q-k)^\mu}{t-q^2}(f_t-\hat{F})
 \nonumber\\
 &\qquad\mbox{}
 -e_f \frac{(2p'-k)^\mu}{u-p'^2}(f_u-\hat{F})
 \nonumber\\
 &\qquad\quad\mbox{}
 -e_i \frac{(2p+k)^\mu}{s-p^2}   (f_s-\hat{F})~,
\end{align}
where
\begin{align}
  \hat{F}= 1-\hat{h}\,\big(1-\delta_s f_s\big) \big(1-\delta_u f_u\big)\big(1-\delta_t
  f_t\big)~.
  \label{eq:Fhatdef}
\end{align}
The factors $\delta_x$ are unity if the corresponding channel contributes to
the reaction in question, and zero otherwise. In principle, the parameter
$\hat{h}$ may be an arbitrary (complex) function, $\hat{h}=\hat{h}(s,u,t)$,
possibly subject to crossing-symmetry constraints.\footnote{Regarding crossing
 symmetry, note that the form (\ref{eq:Fhatdef}) for $\hat{F}$ addresses the concerns raised in
 Ref.~\cite{DW} regarding the original choice for $\hat{F}$ made in
\cite{HH2}.} (However, in the application discussed in the next section, we
simply take $\hat{h}$ as a fit constant. Note that $\hat{h}=0$ corresponds to
Ohta's choice~\cite{Ohta}.) With this choice for $\hat{F}$, the auxiliary
current $C^\mu$ is manifestly
 nonsingular,
\begin{align}
  C^\mu
 &=  -e_\pi(2q-k)^\mu\frac{f_t-1}{t-q^2}\Big[1-\hat{h} \big(1-\delta_s f_s\big)\big(1-\delta_u f_u\big)\Big]
 \nonumber\\[1ex]
 &\quad\mbox{}
 -e_f (2p'-k)^\mu\frac{f_u-1}{u-p'^2}\Big[1-\hat{h}\big(1-\delta_s f_s\big) \big(1-\delta_t f_t\big)\Big]
 \nonumber\\[1ex]
 &\quad\mbox{}
 -e_i (2p+k)^\mu\frac{f_s-1}{s-p^2}  \Big[1-\hat{h}\big(1-\delta_u f_u\big)\big(1-\delta_t f_t\big)\Big]~,
 \label{eq:Cmuexpl}
\end{align}
i.e., it is a \emph{contact} current, and in view of charge conservation,
$e_\pi+e_f-e_i=0$, its four-divergence evaluates to
\begin{equation}
  k_\mu C^\mu = e_\pi f_t + e_f f_u - e_i f_s~.
\end{equation}
With the vertex parametrization (\ref{eq:Fdef}), the gauge-invariance condition
(\ref{eq:McGI}) may now be written explicitly as
\begin{align}
  k_\mu M^\mu_c
  &= g_\pi \gamma_5 k_\mu \Bigg\{
  \left[\lambda +  (1-\lambda) \frac{\fs{q}}{m'+m} \right]  C^\mu
  \nonumber\\[1ex]
  &\qquad\qquad\qquad\mbox{}
  -  (1-\lambda)
\frac{\gamma^\mu}{m'+m}e_\pi  f_t
  \Bigg\}
~,\label{eq:CC1}
\end{align}
or, equivalently, as
\begin{align}
  k_\mu M^\mu_c
  &= g_\pi \gamma_5 k_\mu\Bigg\{
  \left[\lambda +(1-\lambda) \frac{\fs{q}-\fs{k}}{m'+m} \right] C^\mu
  \nonumber\\[1ex]
  &\qquad\qquad\mbox{}
  -(1-\lambda) \frac{\gamma^\mu}{m'+m}\big(e_i f_s- e_f f_u \big)
  \Bigg\}~,
  \label{eq:CC2}
\end{align}
where the respective terms in the braces differ by a manifestly transverse
term. We can exploit this ambiguity and set
\begin{align}
  M^\mu_c
  &= g_\pi \gamma_5 \Bigg\{
  \left[\lambda+  (1-\lambda)\frac{\fs{q}-\beta \fs{k}}{m'+m}
 \right]C^\mu
  \nonumber\\[1ex]
  &\qquad\qquad\mbox{}
 -  (1-\lambda) \frac{\gamma^\mu}{m'+m} \Big[ e_\pi  f_t -\beta k_\rho C^\rho\Big]
  \Bigg\}
  ~,
  \label{eq:MCbeta}
\end{align}
where all terms depending on the free parameter $\beta$ sum up to a transverse
piece $T^\mu_\beta \propto \beta(\gamma^\mu \,k_\rho C^\rho -\fs{k}C^\mu)$. The
parameter $\beta$ then allows us to mix between the pseudovector `$\fs{k}$
content' found in the expressions within the braces on the right-hand sides of
Eqs.~(\ref{eq:CC1}) and (\ref{eq:CC2}), which correspond to $\beta=0$ and
$\beta=1$, respectively.
Note that $\beta=0$ amounts to a `more traditional' treatment of the
Kroll--Ruderman term where the bare $\gamma_5 \gamma^\mu e_\pi$ coupling is
simply dressed by the $t$-channel form factor $e_\pi \to e_\pi f_t$. For
$\beta=1$, by contrast, this dressing occurs via the linear combination $e_\pi
\to e_i f_s-e_f f_u$ of $s$- and $u$-channel form factors (which in general
would be non-zero even for $\pi^0$ production).

Obviously, the above choice of $M^\mu_c$ is not unique, for we can always add
another transverse current to it. In this sense, the pieces proportional to the
parameter $\beta$ in Eq.~(\ref{eq:MCbeta}) is just a particular choice of the
transverse current added to the contact current. We emphasize, however, that
the transverse contact current in $M^\mu_c$ must not be confused with the
transverse contact current $T^\mu$ appearing in Eq.~(\ref{eq:MiaMc}). Note, in
particular, that $M^\mu_c$ (and any of its contributing pieces) does not appear
inside the FSI loop integral, but $T^\mu$ does.

\subsubsection{The transverse contact current $T^\mu$}\label{sec:transcontact}

The most general structure of a transverse contact current in pion
photoproduction\footnote{For simplicity, we consider here only real photons,
but these considerations can easily be extended to electroproduction processes
as well.} can be written as \cite{BKLM}
\begin{equation}
T^\mu = \gamma_5\sum_{j=1}^4 A_j  T^\mu_j \ , \label{transc}
\end{equation}
where
\begin{subequations}\label{transop}
\begin{align}
T^\mu_1 & =  \frac{i}{m}\sigma^{\mu\nu}k_\nu =
\frac{1}{m} \left( \gamma^\mu\fs{k} - k^\mu \right) \ , \\
T^\mu_2 & =  \frac{1}{m^3} \left[ P^\mu \left(2q\cdot k -k^2\right) -
\left(2q-k\right)^\mu P\cdot k \right] \ , \\
T^\mu_3 & =  \frac{1}{m^2} \left(\gamma^\mu q\cdot k - q^\mu\fs{k}\right) \ , \\
T^\mu_4 & = \frac{1}{m^2} \left(\gamma^\mu P\cdot k - P^\mu \fs{k}\right) -
T^\mu_1 \ ,
\end{align}
\end{subequations}
with $P^\mu \equiv (p + p')^\mu/2$. The operators $T^\mu_j$ constitute a
complete set of manifestly transverse operators for real photons. The
coefficients $A_j$ should be free of any singularities in order to ensure that
$T^\mu$ is a genuine contact current. The simplest approximation one can make
for these coefficients is to assume them to be of the form
\begin{equation}
A_j = \frac{a_j}{k_0} \ , \label{transcc}
\end{equation}
with $k_0$ denoting here the photon energy and $a_j$ being dimensionless
constants to be fixed by the data. Notice that the factor $1/k_0$ in this
equation will be canceled by the factor $k_0$ which appears once the matrix
element of the transverse current $T^\mu$ is calculated.

To explore the energy range of where the approximation (\ref{transcc}) may be
expected to produce reasonable results, let us calculate the corresponding
matrix element of $T^\mu$,
\begin{equation}
\hat{T} \equiv \bar{u}_{\vec{p}^{\,\prime}}(\epsilon_\mu T^\mu)  u_{\vec{p}} \
, \label{mrtxT}
\end{equation}
where $\epsilon_\mu$ denotes the photon's polarization vector and $u_{\vec{p}}$
is the nucleon spinor, with three-momentum $\vec{p}$, normalized as
$\bar{u}_{\vec{p}} u_{\vec{p}}=1$. (Note that $u_{\vec{p}}$ here does not
contain the Pauli spinor, i.e., $\hat{T}$ is an operator in spin-1/2 space.) We
have, in the center-of-momentum frame of the system,
\begin{equation}
\hat{T} = F_1\, \vec{\sigma} \cdot \vec{\epsilon}
      + i F_2\, \vec{\epsilon} \cdot \hat{n}_2
      +   F_3\, \vec{\sigma} \cdot \hat{k} \, \vec{\epsilon} \cdot \hat{q}
      +   F_4\, \vec{\sigma} \cdot \hat{q} \, \vec{\epsilon} \cdot \hat{q}
      ~, \label{SSTRUC_1P}
\end{equation}
with $\hat{n}_2 \equiv (\hat{k} \times \hat{q})/|\hat{k} \times \hat{q}|$,
where the hats denote unit three-vectors (i.e., $\hat{v} \equiv \vec{v} /
|\vec{v}|$ for an arbitrary vector $\vec{v}$), and
\begin{subequations}\label{coeffT}
\begin{align}
F_1 & = \alpha_0 + \left(\alpha_1
      + \alpha_2 \frac{|\vec{q}\,|}{m}\cos\theta\right)\frac{|\vec{q}\,|}{m}\cos\theta
       \ , \displaybreak[0] \\[1ex]
F_2 & = \left(\beta_1 + \beta_2 \frac{|\vec{q}\,|}{m}\cos\theta\right)
        \frac{|\vec{q}\,|}{m}\sin\theta \ , \displaybreak[0] \\[1ex]
F_3 & = \left( \delta_1 + \delta_2 \frac{|\vec{q}\,|}{m}\cos\theta \right)
        \frac{|\vec{q}\,|}{m} \ ,  \\[1ex]
F_4 & = \eta_2 \frac{|\vec{q}\,|^2}{m^2} \ ,
\end{align}
\end{subequations}
with $\cos\theta\equiv \hat k \cdot \hat q$. The quantities $\alpha_i$,
$\beta_i$, $\delta_i$, and $\eta_i$ are given explicitly in Appendix A in terms
of the coefficients $a_j$ of Eq.~(\ref{transcc}). These results show explicitly
that the constant approximation for the coefficients $a_j$ of (\ref{transcc})
leads to a transverse contact current $T^\mu$ that accounts for parts of the
partial wave contributions up to $D$ waves in the final $\pi N$
state\footnote{See also Ref.~\cite{NL}, where the coefficients $F_j$ in
Eq.~(\ref{SSTRUC_1P}) are given in terms of the partial-wave matrix elements to
any desired order of the expansion.}. Therefore, the constant approximation
(\ref{transcc}) should be suited for energies not too far from threshold. For
higher energies, if higher partial-wave contributions should be needed, one
might expand $A_j$ in terms of Legendre polynomials of higher order and fit the
corresponding coefficients.

The isospin structure of the transverse contact current can be included
explicitly by writing the coefficient $A_j$ in Eq.~(\ref{transc}) as
\begin{equation}
A_j = \sum_{i=1}^3 \left( A^0_j\tau_i + A^-_j\frac{1}{2}[\tau_i,\tau_3]
    + A^+_j\delta_{i,3} \right) \ ,
\end{equation}
in which case $A^0_j$, $A^-_j$, and $A^+_j$ individually are to be approximated
by Eq.~(\ref{transcc}).

\subsection{Explicitly incorporating the dynamics of exchange currents}

The fitting procedure of $T^\mu$ discussed in the previous subsection provides
an indirect phenomenological means of accounting for the \emph{transverse}
parts of the exchange-current contributions of $U^\mu$ subsumed in
Fig.~\ref{fig:ucurr} which are neglected when approximating $M^\mu_a$ as shown
in Fig.~\ref{fig:MAcurrent}. Specifically, none of the transverse mechanisms
subsumed under the currents $E^\mu$ and $D^\mu$, etc., as defined in
Fig.~\ref{fig:ucurr} \emph{explicitly} enters the approximation procedure
described so far.

This can be done, however, in a systematic order-by-order manner. To this end,
note that with
\begin{equation}
  U^\mu=E^\mu +D^\mu + \cdot
\end{equation}
as defined in Fig.~\ref{fig:ucurr},  we may write Eq.~(\ref{eq:Madef}) as
\begin{equation}
M_a^\mu  = E^\mu G_0 \ket{F\tau} + M'^\mu_a~,
\end{equation}
where
\begin{equation}
M'^\mu_a = m^\mu_\bare +D^\mu G_0 \ket{F\tau} +\cdots~.
 \label{eq:Maprimedef}
\end{equation}
We may now subject $M'^\mu_a$ to the same approximation procedure employed
previously for $M^\mu_a$, now, however, explicitly taking into account the
current $E^\mu$.

One easily finds that \emph{both} Eqs.~(\ref{eq:McGI}) and (\ref{eq:MiaMc})
remain valid, but $T^\mu$ is now given by
\begin{equation}
  T^\mu = (E^\mu - e^\mu ) G_0 \ket{F\tau} +T'^\mu~,
  \label{eq:Tmuexpand}
\end{equation}
where $e^\mu$ is the \emph{non-transverse} part of $E^\mu$ (the same way
$m^\mu_x$ is the non-transverse part of $M^\mu_x$, for $x=u,t$), i.e.,
\begin{equation}
  k_\mu(E^\mu - e^\mu ) G_0 \ket{F\tau} =0~.
\end{equation}
The remaining transverse current $T'^\mu$,
\begin{equation}
k_\mu T'^\mu=0~,
\end{equation}
remains undetermined by this procedure. It plays the same role, obviously, at
the present level that $T^\mu$ played at the previous level of approximation.
We may, therefore, either treat it completely phenomenologically in the manner
of Eq.~(\ref{transc}), or we may, in principle, treat its dynamics explicitly
by now incorporating the triangle-graph currents $D^\mu$ of
Fig.~\ref{fig:ucurr}. This then leads to
\begin{equation}
  T'^\mu = (D^\mu - d^\mu ) G_0 \ket{F\tau} +T''^\mu~,
\end{equation}
where $d^\mu$ is the non-transverse part of $D^\mu$ and $T''^\mu$ the remaining
unspecified transverse contribution.

In principle, one may in this manner include more and more complex dynamical
mechanisms explicitly into the formalism in a step-by-step procedure. In
practice, however, even incorporating the first step, Eq.~(\ref{eq:Tmuexpand}),
explicitly in the interaction current (\ref{eq:MiaMc}) is a highly non-trivial
task since this involves a double-loop integral which is very costly to
evaluate numerically.

Note, however, that as far as gauge invariance is concerned, one need
\emph{not} work with the full currents $E^\mu$ or $D^\mu$, etc. Providing
everything else is done consistently in the manner outlined here, \emph{any}
approximation of, for example, $E^\mu$ that satisfies the gauge-invariance
constraint
\begin{align}
k_\mu E^\mu(p',q',p,q) &= Q'_\pi\, E(p',q'-k,p,q)
   \nonumber\\[1ex]
  &\mbox{}\quad+Q'_N\, E(p'-k,q',p,q)
  \nonumber\\[1ex]
  &\mbox{}\qquad
  - E(p',q',p,q+k)\,Q_\pi
    \nonumber\\[1ex]
  &\mbox{}\qquad\quad
  - E(p',q',p+k,q)\,Q_N
\end{align}
will preserve gauge invariance as a matter of course \cite{HH1}. Here, $E$
describes the simple exchange graph obtained by stripping the photon off any of
the three contributions to $E^\mu$, and $q,q'$ and $p,p'$ are the
incoming/outgoing pion and nucleon momenta, respectively, related by
\begin{equation}
  p+q+k =p'+q'~,
\end{equation}
and $Q_\pi,Q'_\pi$ and $Q_N,Q'_N$ the corresponding charge operators. A similar
equation can be written down for $D^\mu$ and, for that matter, for any
topologically distinct contribution to $U^\mu$.

\subsection{Unitarity}

The full field-theoretical photoproduction formalism as summarized in
Fig.~\ref{diagrams1} is unitary as a matter of course (within the limits of the
usual one-photon approximation). It is a straightforward exercise to show that
the discontinuity contributions for the corresponding unitarity relation arise
from the intermediate $\pi N$ propagator $G_0$ in the hadronic relation
\begin{equation}
 X=U+UG_0X
\end{equation}
for the non-polar $\pi N$ amplitude and from the \emph{dressed} single-nucleon
$s$-channel pole term $\ket{F\tau}S\bra{F\tau}$ that appears in the full $\pi
N$ $T$-matrix,
\begin{equation}
  T=\ket{F\tau}S\bra{F\tau} +X~.
\end{equation}
(The preceding two equations are depicted in the last two lines of diagrams in
Fig.~\ref{diagrams1}(b); see also Ref.~\cite{HH1}.) In the one-photon limit,
therefore, only the cut structure of the \emph{hadronic} part of the
photo-reaction is relevant and other cuts do \emph{not} contribute to the
unitarity relation. This applies, in particular, to the cuts that are contained
in the contact current $M^\mu_a$ shown in Fig.~\ref{fig:MAcurrent}. Hence,
\emph{any} approximation of $M^\mu_a$ will preserve the unitarity of the
photoproduction amplitude. The approximation choice (\ref{eq:Ma_approx}),
therefore, does not violate unitarity.

\section{Application: Pion Photoproduction}

In this section, we apply the approach developed in the preceding section to
the photoproduction reaction $\gamma + N \to \pi + N$. At this stage, this is
intended to be more of a feasibility study rather than a serious attempt at
describing all features of the data. We will therefore make some simplifying
assumptions along the way.

In the present application, we restrict ourselves to photon energies up to
about $400$ MeV. Therefore, in addition to the basic nucleons and pions
discussed in the preceding section, our model also incorporates intermediate
$\Delta$'s in the $s$- and $u$-channels. Details of the dressing of the
electromagnetic $N\Delta$ transition vertex in the present approach is given in
Appendix B. We also include the $\rho$, $\omega$, and $a_1$ meson exchanges in
the $t$-channel. Note here that transition currents between different hadronic
states are transverse individually and therefore play no role for the issue of
gauge invariance. The details of the respective interactions are specified by
the Lagrangian densities given in Appendix C. Form factors are attached to the
hadronic vertices to account for the off-shellness of the respective
intermediate hadrons. The details of these form factors are also found in
Appendix C.

For the $\pi N$ FSI, we employ the $\pi N$ $T$-matrix developed by the J\"ulich
group \cite{Oliver} which results from a dynamical model based on a
coupled-channels approach. Among other things, this interaction fits the $\pi
N$ phase-shifts and inelasticities below about 1.5 GeV. For larger energies it
provides a background due to final state interactions which has to be
supplemented by baryon resonances. It should be noted that the J\"ulich $\pi N$
interaction is based on time-ordered perturbation theory (TOPT)
\cite{schweber}. Therefore, to be fully consistent with this interaction, one
should also evaluate the amplitudes $M^\mu_x$ ($x=s,u,t$) within TOPT. In the
present application, however, we have ignored this consistency requirement and
evaluate these amplitudes following the Feynman prescription (which coincides
with TOPT at the tree level). As a consequence, there is an ambiguity in
defining the zeroth component of the four momentum of the intermediate state in
the transverse amplitudes $(M^\mu_u-m^\mu_u)$ and $(M^\mu_t-m^\mu_t)$ appearing
under the loop integral of the FSI contribution in Eq.~(\ref{eq:MiaMc}). We
follow the choice made (based on the gauge invariance consideration) in
Ref.~\cite{Nozawa} for the zeroth component of the intermediate particle
momentum in evaluating these amplitudes.\footnote{Note that, in the present
approach, the photoproduction amplitude is gauge invariant independent of this
particular choice of defining the zeroth component of the intermediate particle
momentum.}

In the present application, for simplicity, we ignore the dressing effects in
the $s$-channel nucleon pole propagator. Moreover, we ignore the explicit
dressing of the nucleon electromagnetic vertex as given by
Eq.~(\ref{dvrtx_approx}).
 Instead, we take the vertex given by
Eq.~(\ref{LagNDgama}) with \emph{physical} coupling constants. The analogous
approximation is also adopted for the $N\Delta\gamma$ vertex given by
Eq.~(\ref{LagNDgamg}). Here, the coupling constants $G_1$ and $G_2$ are treated
as free parameters adjusted to reproduce the data. Such an approximation is not
critical for the present purpose of illustrating how the method developed in
the previous section works. Obviously, dressing of the electromagnetic vertices
is more critical in electroproduction processes.

\begin{table}[t!]
\begin{center}
\caption{Model parameters fitted to the reaction $\gamma N \to  \pi N$. The
$N\Delta\gamma$ coupling constants $G_1$ and $G_2$ are constrained by the
measured E2 to M1 ratio of $R_{EM} = -2.5\% $ \cite{Mainz1,Mainz2}. The
parameter $\beta$ in the contact current is fixed to be zero from the outset.
Moreover, we consider only pure pseudovector coupling and therefore always have
$\lambda=0$.}
\begin{tabular}{@{\qquad}c@{\qquad}c@{\qquad}c@{\qquad}c@{\qquad}}
\hline\hline
   $G_1$    &   $G_2$   &   $\Lambda$ (MeV)   &   $\hat h$        \\
\hline
   $3.84$   &   $-1.94$ &   $604$             &   $0.01$          \\
\hline
\end{tabular}
\label{tab:FREEPAR}
\end{center}
\end{table}
%

In the following, we list the free parameters of the model in the present
application.
\begin{itemize}
\item[1)]
$G_1$ and $G_2$: The two dressed electromagnetic coupling constants at the
$N\Delta\gamma$ vertex, as specified in Appendix C, are not independent of each
other, for we impose the E2 to M1 ratio to be $R_{EM}=-2.5\% $ as determined by
the Mainz group \cite{Mainz1,Mainz2}.
\item[2)]
$\hat h$,  $\beta$, and $\lambda$: The parameters $\hat h$ and $\beta$ appear
in the contact current in Eqs.~(\ref{eq:Cmuexpl}) and (\ref{eq:MCbeta}), and
the PS/PV mixing parameter $\lambda$ appears in Eq.~(\ref{eq:Fdef}). We take
$\beta=0$ and $\lambda=0$ (pure PV) from the outset, so that $\hat h$ is the
only parameter to be fixed in the contact current $M^\mu_c$.
\item[3)]
$T^\mu$: The transverse contact current in Eq.~(\ref{eq:MiaMc}) is found to be
negligible in the present application, i.e., the parameters $a_j$ in
Eq.~(\ref{transcc}) are taken as $a_j = 0$, $(j=1,2,3,4)$. In other words, we
found no need for such a current in reproducing the cross-section data.
\item[4)]
 $\Lambda$: This regularization parameter is needed for the loop integral in the FSI
contribution. We regularize this integral by introducing a cutoff function
\begin{equation}
F_R = \frac{\Lambda^2}{\Lambda^2 + \vec{q}\ ^2} \ , \label{regul}
\end{equation}
where $\vec{q}$ denotes the relative momentum of the intermediate $\pi N$ state
in the loop integral. Of course, this regulator may also be interpreted as the
form factor which accounts for the off-shellness of the pion and nucleon in the
loop integral.
\end{itemize}
%

\begin{figure}[t!]\centering
\includegraphics[height=\columnwidth,angle=-90,clip]{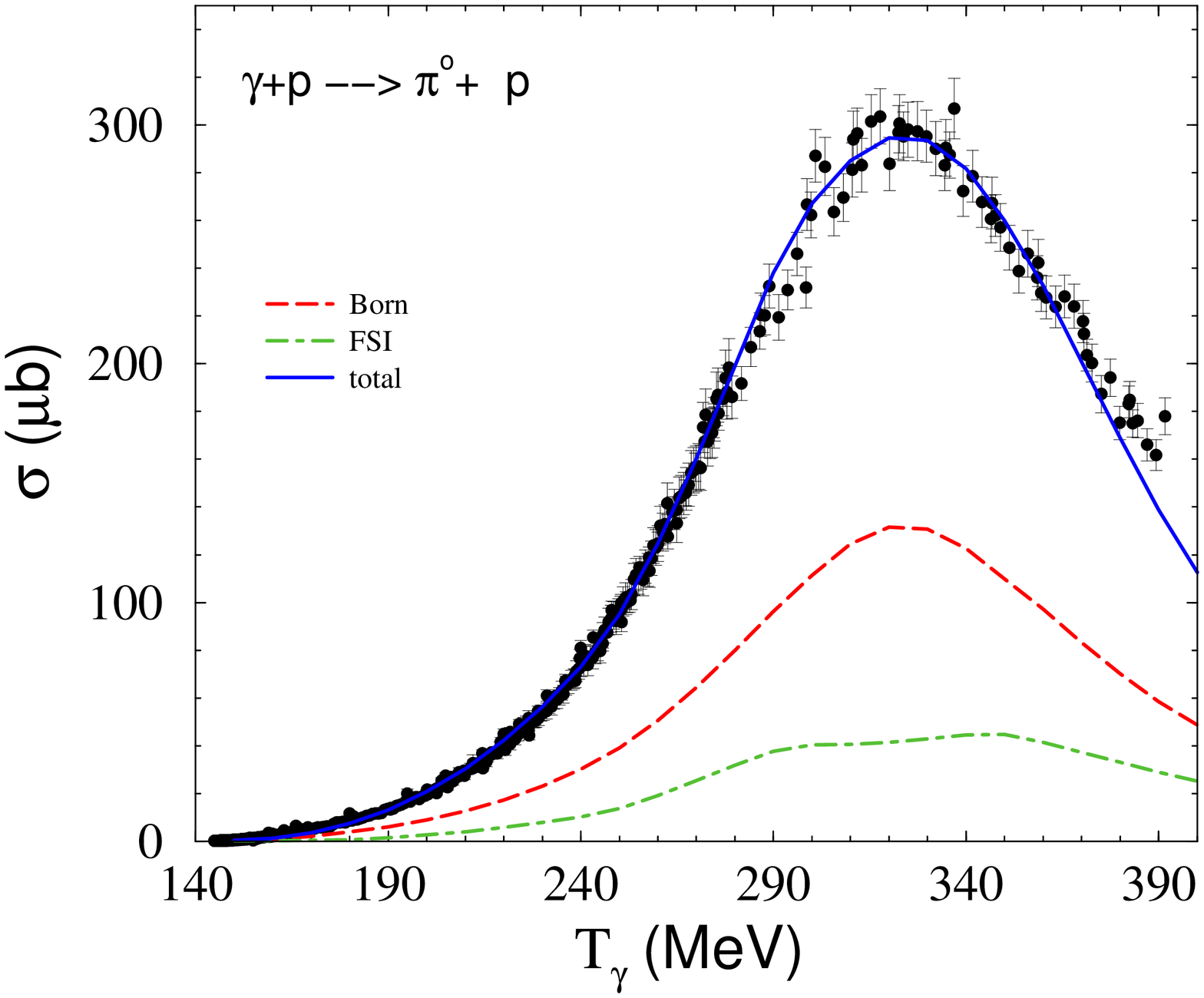}
\caption{\label{fig:txsc}%
(Color online) Results corresponding to the parameter set in Table~I for the
total cross section as a function of photon incident energy $T_\gamma$ in the
reaction $\gamma + p \rightarrow \pi^0 + p$. The dashed curve corresponds to
the Born contribution and the dash-dotted curve to the FSI loop contribution.
The solid curve is the total contribution. The data are from Ref.~\cite{txsc}.}
\end{figure}
%

\begin{figure}[b!]\centering
\includegraphics[width=\columnwidth,angle=0,clip]{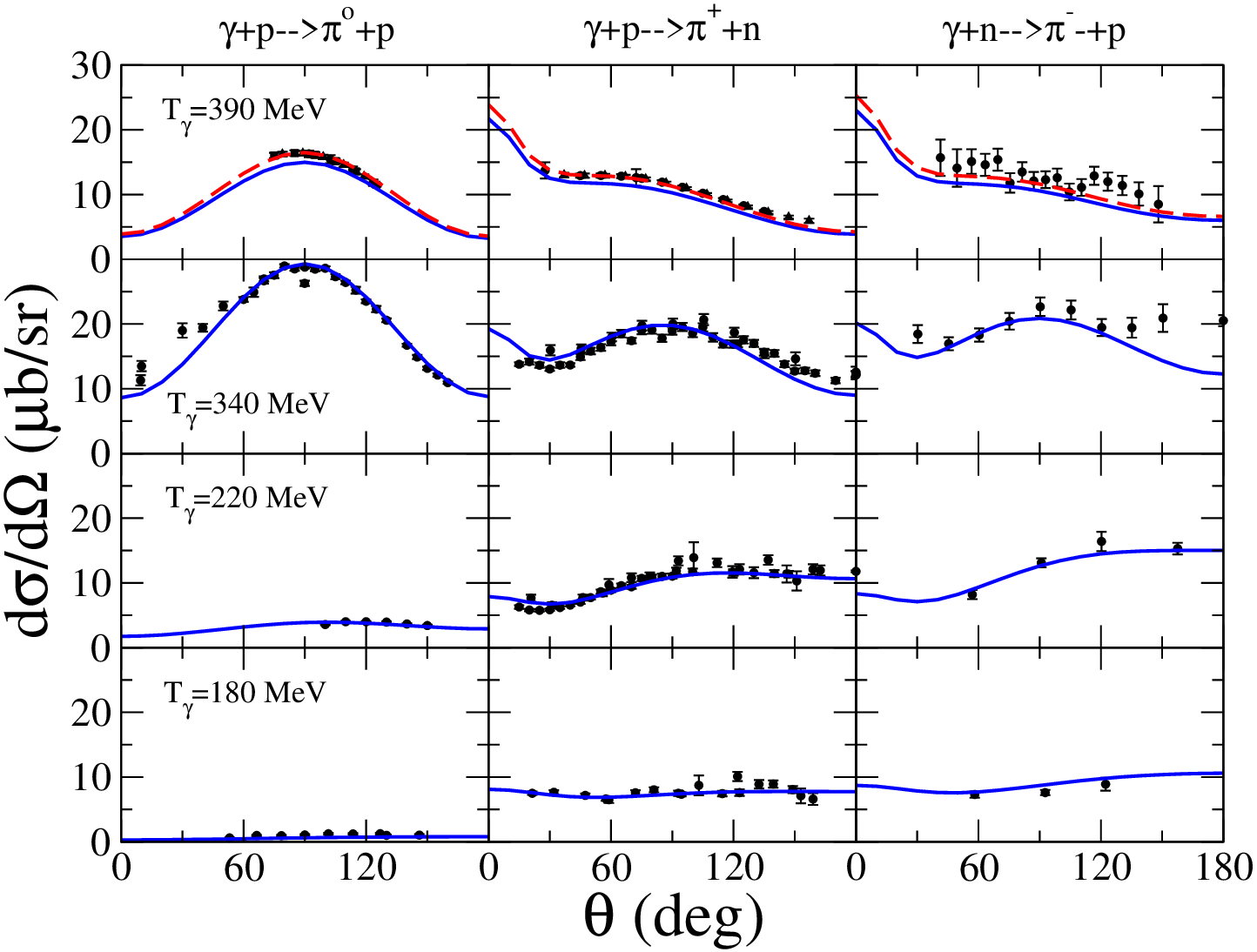}
\caption{\label{fig:dxsc}%
(Color online) Results corresponding to the parameter set in Table~I for the
differential cross sections in the c.m.\ frame of the system in the reaction
$\gamma + N \rightarrow \pi + N$ at various photon incident energies
$T_\gamma$. In the top row, the dashed curves correspond to the results
represented by solid curves multiplied by an arbitrary factor of 1.1 The data
are from  Refs.~\cite{Mainz2,dxsc1}.}
\end{figure}
%

With the considerations mentioned above, we are left with only three
independent free parameters in the present model. They are adjusted to
reproduce the pion photoproduction cross-section data. The resulting parameter
values are given in Table~\ref{tab:FREEPAR}. Note that $\hat h$ is nearly zero,
corresponding practically to the Ohta's choice \cite{Ohta}.

Figure~\ref{fig:txsc} shows the total cross-section result for the reaction
$\gamma + p \to \pi^0 + p $ from the threshold up to $T_\gamma \approx 400$
MeV. As we can see, the agreement with the data is very good except for
energies above $T_\gamma \sim 360$ MeV, where the prediction tends to
underestimate the data. In particular, around $T_\gamma \approx 390$ MeV, the
discrepancy is about $10\%$. We also see that the FSI loop contribution is
relatively small compared to the Born contribution. However, it plays a crucial
role in reproducing the observed energy dependence through its interference
with the dominant Born term.

Figure~\ref{fig:dxsc} shows the results for differential cross sections for
neutral and charged pion productions at various energies together with the
data. We see that, overall, the data are reproduced quite well. The dashed
curves in the top row correspond to the results represented by the solid curves
multiplied by an arbitray factor of 1.1. They are shown here to facilitate
visualizing that the shape of the angular distribution is well reproduced, in
spite of the absolute normalization being underestimated at this energy by
$\sim 10\%$, as can be seen better in Fig.~\ref{fig:txsc}.

\section{Summary}

By exploiting the generalized Ward-Takahashi identity for the production
amplitude and total charge conservation, we have constructed a fully
gauge-invariant (pseudoscalar) meson photoproduction amplitude which includes
the hadronic final-state interaction explicitly. The method is based on a
field-theoretical approach developed earlier by Haberzettl \cite{HH1}. It is
quite general and can be readily extended to any other meson photo- and
electroproduction reactions. This method should be particularly relevant for
the latter reaction.

As an example of application of the present approach, we have calculated both
the neutral and charged pion photoproduction processes off nucleons up to about
$400$ MeV photon incident energy which illustrates the feasibility of the
present method.

Obviously, for a more quantitative calculation, including not only cross
sections but also other observables, some of the approximations made in the
present feasibility study should either be improved or altogether avoided. In
particular, the dressing of the electromagnetic vertices as given by
Eq.~(\ref{dvrtx}) needs to be carried out. Also, for pion photoproduction, one
should constrain the parameters of the present model, if possible at all, by
comparing the amplitude of the present approach in the chiral limit with that
of the Chiral Perturbation Theory. Work in this direction is in progress.

\begin{acknowledgments}
This work was supported by the COSY Grant No.\ 41445282\,(COSY-58).
\end{acknowledgments}

\appendix

\section{Transverse contact current}

In this appendix, we give the explicit formulas for the coefficients
$\alpha_i$, $\beta_i$, $\delta_i$, and $\eta_i$ appearing in
Eq.~(\ref{coeffT}). They are
\begin{subequations}
\begin{align}
\alpha_0 & = N \left[a_1G + a_3\frac{\omega_q}{m}
           + a_4\left(\frac{I}{m} - 2G\right)\right] \ , \\
\alpha_1 & = N \left[- a_1JG - a_3\left(1+\omega_q H\right)
           + a_4\left(1 - IH - 2JG\right) \right] \ ,  \\
\alpha_2 & = N \left[a_3 - a_4 \right]mH \ , \label{alpha}
\end{align}
\end{subequations}%
where $ \omega_q \equiv \sqrt{\vec{q}^{\;2} + m_\pi^2}$ and
\begin{subequations}
\begin{align}
N & \equiv \sqrt{\frac{\varepsilon_q+m}{2m}}\sqrt{\frac{\varepsilon_k+m}{2m}}\frac{1}{m} \ , \\
G & \equiv 1+\frac{|\vec{k}|}{\varepsilon_k+m} \ , \displaybreak[0]\\
H & \equiv  \frac{|\vec{k}|}{\left(\varepsilon_q+m\right)\left(\varepsilon_k+m\right)} \ ,\displaybreak[0]\\
I & \equiv \varepsilon_k+|\vec{k}|+\varepsilon_q \ ,  \\
J & \equiv \frac{m}{\varepsilon_q+m} \ ,
\end{align}
\end{subequations}
with $\varepsilon_p \equiv \sqrt{\vec{p}^{\;2} + m^2}$. Furthermore,
\begin{subequations}
\begin{align}
\beta_1 & = N \left[ -a_1JG + a_3\omega_q H + a_4\left(IH + 2JG\right) \right] \ ,
\\[1ex]
\beta_2 & = - \alpha_2 \ ,
\end{align}
\end{subequations}
and
\begin{subequations}
\begin{align}
\delta_1 & = N \Big[ -a_1JG - a_2\frac{|\vec{k}|}{m}K + a_3\left(G+\omega_q H\right) \\[.5ex]
         &\qquad\mbox{}
         + a_4\left(G - IH - 2JG\right) \Big] \ , \nonumber \\[1ex]
\delta_2 & = - N \left[a_3 + a_4 \right]mH \ ,
\end{align}
\end{subequations}
with $K \equiv 2(\varepsilon_k+m)/(\varepsilon_q+m)$, and finally
\begin{equation}
\eta_2  = N \left[ a_2K - \left(a_3 -
            a_4\frac{|\vec{k}|}{m}\right)\left(J - m H \right) \right] \ .
\end{equation}
%

\section{\boldmath Transversality of the dressed $N\Delta\gamma$ vertex}

For completeness, in this appendix we show how to dress the $N\Delta\gamma$
transition vertex in the present approach and demonstrate that the resulting
vertex is purely transverse as it should be.

Bare transition currents can easily be made transverse by expanding the current
in an appropriate transverse operator base. It is not obvious, however, that
the transversality will remain true after one dresses the current. It will be
shown here that this is indeed the case. As a specific example, we will treat
the electromagnetic current for the $N\to \Delta$ transition.

\begin{figure}[b!]
\parbox{\columnwidth}{\centering
  \includegraphics[width=.9\columnwidth,clip=]{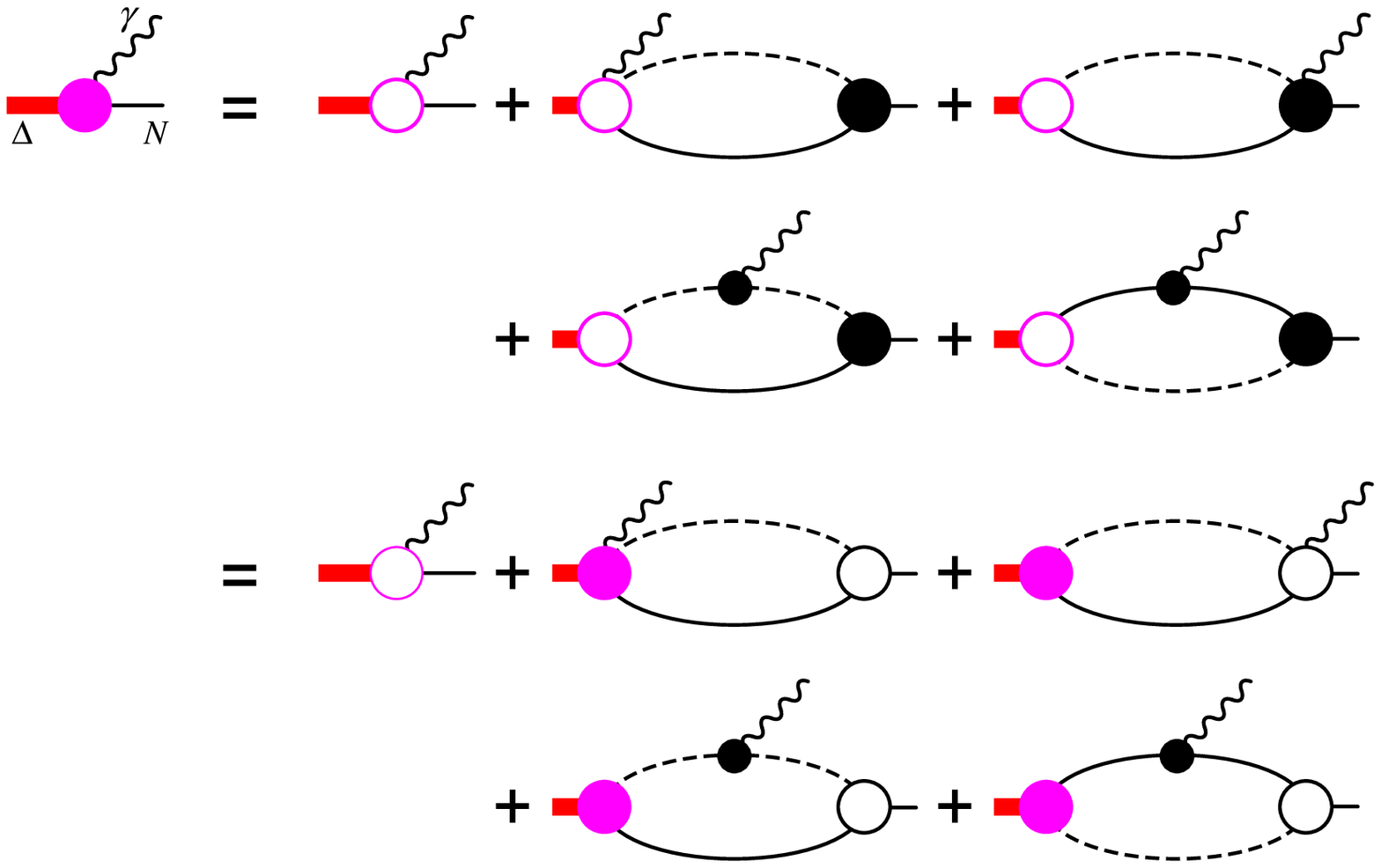}}
  \hfill
  \parbox{\columnwidth}{%
  \caption{\label{fig:gND}%
  (Color online) Dressed transition current for $\gamma N \to \Delta$. Solid circles
  depict dressed vertices and currents, whereas open circles
  show the corresponding bare quantities.
  The two equivalent forms differ by whether the initial $\pi NN$ vertex or the
  final $\pi N\Delta$ vertex is fully dressed. Contributions with intermediate particles
  other than pions and nucleons are not shown.}}
\end{figure}

\begin{figure}[b]
\parbox{\columnwidth}{\centering
  \includegraphics[width=.9\columnwidth,clip=]{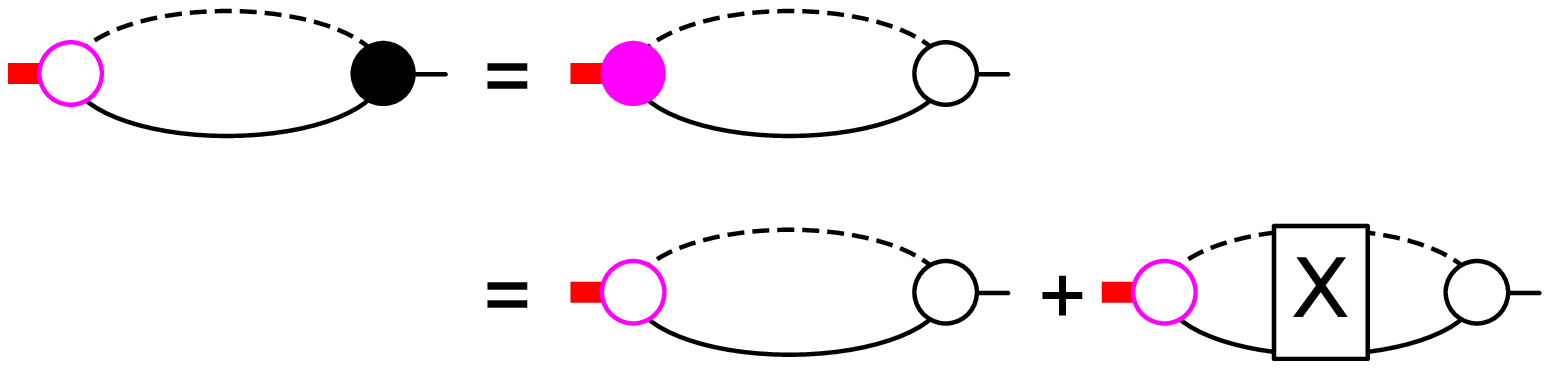}}
  \hfill
  \parbox{\columnwidth}{%
  \caption{\label{fig:NDbubble}%
  (Color online) Topological structure of the bubbles underlying the construction of the
  transition current in Fig.~\ref{fig:gND}. The first equality follows from the
  fact that the dressing for both bubbles is done in terms of the non-polar
  $\pi N$ $T$-matrix $X$ and both bubbles with dressed vertices, therefore, can
  be written in terms of the graphs of the second line. Taken as a physical
  process, the transition $N\to\Delta$ is not possible either on
  or off-shell because of isospin conservation. }}
\end{figure}

The dressing mechanism for this current is depicted in Fig.~\ref{fig:gND} which
is constructed in analogy to the dressed nucleon current in (\ref{dvrtx}) (see
also Ref.~\cite{HH1} for full detail). We will consider here only dressing
mechanisms in terms of nucleons and pions. Other particles are independent from
the ones considered here and formally would not add anything new except
complicating the presentation. The two equivalent forms arise from attaching
the photon in all possible ways to two equivalent bubbles shown in
Fig.~\ref{fig:NDbubble}. It should be obvious, however, that \emph{as a
physical process} the transition $N\to\Delta$ as shown in this figure is not
possible because of isospin conservation. Within the present context,
therefore, the bubbles in Fig.~\ref{fig:NDbubble} form but the
\emph{topological} backdrop against which the current shown in
Fig.~\ref{fig:gND} is constructed but are not considered as having a physical
meaning of their own.

Following Ref.~\cite{js72}, the most general transverse current for
\begin{equation}
\gamma(k)+N(p) \to \Delta(p')
\end{equation}
may be written as
\begin{align}
\Gamma^{\beta\mu} & = G_1 \,\gamma_5\left(k^\beta \gamma^\mu-g^{\beta\mu}\,
\fs{k}\right)
\nonumber\\
&\qquad\mbox{} +G_2\,\gamma_5\left(k^\beta
P^\mu-g^{\beta\mu} \,\sdot{k}{P}\right)
\nonumber \\
&\qquad\quad\mbox{}
+  G_3\,\gamma_5 \left(k^\beta k^\mu-
g^{\beta\mu}\,k^2\right)~, \label{eq:transansatz}
\end{align}
where $P=(p+p')/2=(2p+k)/2$; the Lorentz indices $\mu$ and $\beta$ pertain to
the photon and the $\Delta$, respectively. The $G_i$ are the corresponding form
factors; for real photons, only $G_1$ and $G_2$ contribute. This is the ansatz
that one chooses for the bare current.

To show that the transversality of the bare current is preserved if one now
dresses the current, we use the first form of the current shown in
Fig.~\ref{fig:gND}. Using the notation of Ref.~\cite{HH1}, we can translate
this into the schematic equation
\begin{align}
\Gamma^{\beta\mu}&=\Gamma_0^{\beta\mu}
+\bra{f^{\beta\mu}}S_N\circ\Delta_\pi\ket{F}
 \nonumber\\[1ex]
 &\qquad\quad\mbox{}
 +\bra{f^{\beta}}S_N\circ\Delta_\pi\ket{M^\mu_{\text{int}}}
 \nonumber\\[1ex]
 &\qquad\quad\mbox{}
 +\bra{f^{\beta}}S_N\circ[\Delta_\pi \Gamma^\mu_\pi\Delta_\pi]\ket{F}
 \nonumber\\[1ex]
 &\qquad\quad\mbox{}
 +\bra{f^{\beta}}[S_N\Gamma^\mu_N S_N]\circ \Delta_\pi\ket{F}~,
\end{align}
where the order of terms is exactly as in Fig.~\ref{fig:gND}. The (transverse)
bare current is denoted by $\Gamma_0^{\beta\mu}$; $f^\beta$ is the bare $\pi
\Delta N$ vertex and $f^{\beta\mu}$ the corresponding bare contact current. The
latter are given by
\begin{equation}
f^\beta = f_\Delta\, \theta^{\beta\nu} q_\nu~,
\end{equation}
where $q_\nu$ is the incoming pion momentum, $\theta^{\beta\nu}$ the coupling
tensor, and $f_\Delta$ the bare coupling constant, and by
\begin{equation}
f^{\beta\mu} = -f_\Delta\, \theta^{\beta\mu} e_\pi~,
\end{equation}
where $e_\pi$ is the charge of the intermediate pion. $F$ is the dressed $\pi
NN$ vertex and $M^\mu_{\text{int}}$ the corresponding dressed interaction
current. The convolution of the intermediate pion and nucleon propagators (with
and without attached photons) is denoted by $A\circ B$. The momentum dependence
is suppressed here, but can easily be found by noting that the initial nucleon
momentum is $p$ and that the photon feeds a momentum $k$ into the line (or
vertex) to which it is attached.

Using the Ward--Takahashi identities for the pion, the nucleon, and the
interaction current,
\begin{subequations}
 \begin{align}
 k_\mu S_N \Gamma^\mu_N S_N &= e'_N \left(\kinject S_N-S_N
 \kinject\right)~,
\\[2ex]
 k_\mu \Delta_\pi \Gamma^\mu_\pi \Delta_\pi &= e_\pi \left(\kinject \Delta_\pi-\Delta_\pi
 \kinject\right)~,
\\[2ex]
k_\mu M^\mu_{\text{int}} &= -F [\kinject e_N]  + [e'_N \kinject] F +[e_\pi
\kinject] F~,
 \end{align}
\end{subequations}
respectively, where the solid dot ($\kinject$) indicates at which point the
photon momentum is injected into the equations. For example,
\begin{equation}
e'_N \left(\kinject S_N-S_N
 \kinject\right) = e'_N \Big[S_N(p'')-S_N(p''+k)\Big]~,
\end{equation}
where $p''$ is the initial momentum  of the nucleon with charge $e'_N$ within
the loop. The notation $[e_x \kinject]$ specifies that the photon momentum is
injected into the (incoming or outgoing) particle line with charge $e_x$. We
now find
\begin{align}
k_\mu \Gamma^{\beta\mu} &= \underbrace{k_\mu \Gamma_0^{\beta\mu}}_{=0}
-f_\Delta e_\pi \bra{\theta^{\beta\mu}k_\mu}S_N\circ\Delta_\pi\ket{F}
 \nonumber\\[1ex]
 &\qquad\mbox{}
  - e_N\bra{f^\beta}S_N\circ\Delta_\pi\ket{F} \kinject
 \nonumber\\[1ex]
 &\qquad\mbox{}
 + e_\pi\bra{f^\beta}S_N\circ [\Delta_\pi \kinject]\ket{F}
  \nonumber\\[1ex]
 &\qquad\mbox{}
  + e'_N\bra{f^\beta }[S_N\kinject]\circ\Delta_\pi\ket{F}
   \nonumber\\[1ex]
 &\qquad\mbox{}
  + e_\pi\bra{f^\beta}S_N\circ[\kinject\Delta_\pi] \ket{F}
 \nonumber\displaybreak[0]\\[1ex]
 &\qquad\mbox{}
- e_\pi\bra{f^\beta}S_N\circ[\Delta_\pi\kinject]\ket{F}
 \nonumber\\[1ex]
 &\qquad\mbox{}
 +e'_N\bra{f^{\beta}}[\kinject S_N]\circ \Delta_\pi\ket{F}
 \nonumber\\[1ex]
 &\qquad\mbox{}
 -e'_N\bra{f^{\beta}}[S_N \kinject ]\circ \Delta_\pi\ket{F}
 \nonumber\displaybreak[0]\\[1ex]
&=  -f_\Delta e_\pi \bra{\theta^{\beta\mu}k_\mu}S_N\circ\Delta_\pi\ket{F}
 \nonumber\\[1ex]
 &\qquad\mbox{}
 - e_N\bra{f^\beta}S_N\circ\Delta_\pi\ket{F} \kinject
 \nonumber\\[1ex]
 &\qquad\mbox{}
 + e_\pi\bra{f^\beta }S_N\circ[\kinject\Delta_\pi] \ket{F}
 \nonumber\\[1ex]
 &\qquad\mbox{}
 +e'_N\bra{f^{\beta}}[\kinject S_N]\circ \Delta_\pi\ket{F} ~,
\end{align}
In the second term, the dot $\kinject$ simply indicates that the overall
four-momentum of $\bra{f^\beta}S_N\circ\Delta_\pi\ket{F}$ is $p+k$. In the
first, third, and fourth terms, the intermediate pion propagator depends on the
same loop variable $q$ in all terms, but the left-most vertices have momentum
dependencies that can be combined according to
\begin{align}
 -f_\Delta& e_\pi \bra{\theta^{\beta\mu}k_\mu}S_N\circ\Delta_\pi\ket{F}
 + e_\pi\bra{f^\beta }S_N\circ[\kinject\Delta_\pi] \ket{F} & \nonumber\\[1ex] &
+e'_N\bra{f^{\beta}}[\kinject S_N]\circ \Delta_\pi\ket{F} &
\nonumber\\[1ex]
&\quad= f_\Delta \Big[-e_\pi \bra{\theta^{\beta\mu}k_\mu}+e_\pi
\bra{\theta^{\beta\nu}(q+k)_\nu} \nonumber\\[1ex]
 &\qquad\qquad\qquad\qquad\mbox{}
  +e'_N
\bra{\theta^{\beta\nu}q_\nu}\Big]S_N\circ\Delta_\pi \ket{F}
\nonumber\\[1ex]
 &\quad = f_\Delta \left[(e'_N+e_\pi)
\bra{\theta^{\beta\nu}q_\nu}\right]S_N\circ\Delta_\pi \ket{F}
\nonumber\\[1ex]
&\quad = (e'_N+e_\pi)
\bra{f^\beta}S_N\circ\Delta_\pi \ket{F} \nonumber\\[1ex]
 &\quad
 \mbox{} = e_N\bra{f^\beta}S_N\circ\Delta_\pi \ket{F}~,
\end{align}
where charge conservation,
\begin{equation}
e_N=e_\pi +e'_N~,
\end{equation}
was used. Comparison with the first bubble of Fig.~\ref{fig:NDbubble} shows
that
\begin{equation}
\Sigma^\beta_{\Delta N}=\bra{f^\beta}S_N\circ\Delta_\pi \ket{F}
\end{equation}
is just equal to this topological bubble. Hence we have
\begin{equation}
k_\mu \Gamma^{\beta\mu} = e_N\Big[\Sigma^\beta_{\Delta
N}(p)-\Sigma^\beta_{\Delta N}(p+k)\Big]~.
\end{equation}
As explained above, the $\Sigma^\beta_{\Delta N}$ do not describe a physical
process and vanish individually. Hence, we have
\begin{equation}
k_\mu \Gamma^{\beta\mu} = 0
\end{equation}
and the dressed current thus is also transverse.

We observe that the present approach for dressing the $N\Delta\gamma$ vertex
differs from the dressing mechanism employed in other approaches (see, e.g.,
\cite{Sato,Pascalutsa}) by the presence of the second diagram on the
right-hand-side of the equality in Fig.~\ref{fig:gND}. Note that this term
involves the three-particle to one-particle transition, $\gamma\pi N \to
\Delta$, and therefore is outside the model space considered in those
approaches. We emphasize, however, that the presence of the $\Delta\pi N\gamma$
contact vertex is absolutely necessary for preserving the gauge invariance of
the dressed vertex in view of the momentum dependence of the $N\Delta\pi$
vertex as given by Eq.~(\ref{LagNDpi}) below. This suggests, of course, that
indiscriminate truncation of the model space along particle numbers is not a
good dynamical ordering scheme as far as the gauge-invariance condition is
concerned.

\section{Interactions}

Our model for the $s$-, $u$-, and $t$-channel amplitudes $M^\mu_s$, $M^\mu_u$
and $M^\mu_t$, respectively, in Eq.~(\ref{eq:Mmu_suti}) is constructed from the
interaction Lagrangian density written as a sum of two terms, ${\cal L}_\ia =
{\cal L}_{\text{hadr}} + {\cal L}_{\text{elec}}$, where ${\cal
L}_{\text{hadr}}$ denotes the part of the interaction Lagrangian containing
only the hadron fields, and ${\cal L}_{\text{elec}}$ contains the
electromagnetic interaction with hadrons. For ${\cal L}_{\text{hadr}}$, we have
\begin{equation}
{\cal L}_{\text{hadr}} = {\cal L}_{NN\pi}
                + {\cal L}_{NN\rho}+ {\cal L}_{NNa_1} + {\cal L}_{NN\omega}
                + {\cal L}_{N\Delta\pi} \ ,
\label{Laghadr}
\end{equation}
with
\begin{subequations}
\begin{align}
{\cal L}_{NN\pi} & = - g_\pi \bar\Psi \left( \gamma_5 \left[ i\lambda + \frac{1
- \lambda}{m+m'}\, \fs{\partial} \right]\vec{\pi}\cdot\vec\tau \right)
 \Psi ~,\\[1ex]
 {\cal L}_{NN\rho}& =  - \frac{g_\rho}{2}\; \bar\Psi
\left(\gamma_\mu - \frac{\kappa_\rho}{2m_N}\sigma_{\mu\nu}\partial^\nu \right)
   \vec{\tau} \cdot \vec{\rho}^{\,\mu}  \; \Psi ~,\displaybreak[0]\\[1ex]
 {\cal L}_{NN\omega} & =  - \frac{g_\omega}{2}\; \bar\Psi
\left(\gamma_\mu - \frac{\kappa_\omega}{2m_N}\sigma_{\mu\nu}\partial^\nu\right)
  \omega^\mu  \;  \Psi ~,\displaybreak[0]\\[1ex]
 {\cal L}_{NNa_1} & =   g_{a_1}\, \bar\Psi\,
   \gamma_\mu\gamma_5\vec{\tau} \cdot \vec{a}_1^\mu\, \Psi ~,\\[1ex]
 {\cal L}_{N\Delta\pi} & =  \frac{f_{N\Delta\pi}}{m_\pi}\,
 \bar{\Psi}_\Delta^\mu\, \vec{T}^\dagger \cdot (\partial_\mu \vec{\pi})\,
  \Psi \ + \  \hc
\label{LagNDpi}
\end{align}\end{subequations}
Here, $\Psi$ and $\Psi_\Delta^\mu$ denote the nucleon and $\Delta$ fields,
respectively, $\vec{\pi}$ the pion field, $\vec{\rho}^{\,\mu}$ the
$\rho$-meson, $\omega^\mu$ the $\omega$-meson, and $\vec{a}_1^\mu$ the
$a_1$-meson fields. The latter is included as the chiral partner of the
$\rho$-meson. The vector notation refers to the isospin space.
$\vec{T}^\dagger$ stands for the isospin 1/2 to 3/2 transition operator. The
nucleon and pion masses are denoted by $m_N$ and $m_\pi$, respectively. The
$g_\pi$, $g_\rho$($\kappa_\rho$), $g_{a_1}$, $g_\omega$($\kappa_\omega$) and
$f_{N\Delta\pi}$ are the corresponding coupling constants. We use $g_\pi=14.4$
and $f_{N\Delta\pi}=0.36$ \cite{Oliver}. For the $NN\rho$ coupling constants,
we use $(g_\rho/2)^2/4\pi = 0.91$ and $\kappa_\rho = 6.1$ \cite{Machleidt},
while $(g_\omega/2)^2/4\pi = 11$ and $\kappa_\omega = 0$ \cite{Janssen}. The
coupling constant $g_{a_1}=m_{a_1}f_\pi/m_\pi$, with $m_{a_1}\approx 1260$ MeV
denoting the mass of the $a_1$-meson, has been fixed from the chiral-symmetry
considerations following the work of Wess and Zumino \cite{Wess}.

The electromagnetic interaction Lagrangian density is given by
\begin{align}
{\cal L}_{\text{elec}} & = {\cal L}_{NN\gamma} + {\cal L}_{NN\pi\gamma} +
                  {\cal L}_{\pi\pi\gamma} + {\cal L}_{\omega\pi\gamma}
                  \nonumber \\
                  &\qquad\mbox{}  +                   {\cal L}_{\rho\pi\gamma}
  +  {\cal L}_{a_1\pi\gamma}  +  {\cal L}_{N\Delta\gamma}~,
  \label{Lagelec}
\end{align}
with
\begin{subequations}
\begin{align}
 {\cal L}_{NN\gamma}   &= - e\, \bar\Psi \left(\hat e \gamma_\mu -
\frac{\hat\kappa}{2m_N}
 \sigma_{\mu\nu}\partial^\nu \right) A^\mu \; \Psi
 \label{LagNDgama} ~\\[1ex]
 {\cal L}_{NN\pi\gamma}  &=  e\, \frac{f_\pi}{m_\pi} \,
 \bar\Psi \gamma_5\gamma_\mu [\vec\tau \times \vec\pi]_3\,
                             \Psi  A^\mu
 ~,\\[1ex]
 {\cal L}_{\pi\pi\gamma}  &=
                    e\, [(\partial_\mu \vec\pi) \times \vec\pi]_3 A^\mu
 ~,\displaybreak[0]\\[1ex]
 {\cal L}_{\omega\pi\gamma}  &= e\,
\frac{g_{\omega\pi\gamma}}{m_\pi} \varepsilon_{\alpha\mu\lambda\nu}\,
(\partial^\alpha A^\mu) (\partial^\lambda \pi_3) \omega^\nu
 ~,\displaybreak[0]\\[1ex]
 {\cal L}_{\rho\pi\gamma}  &=  e\,  \frac{g_{\rho\pi\gamma}}{m_\pi}
 \varepsilon_{\alpha\mu\lambda\nu}\, (\partial^\alpha A^\mu) (\partial^\lambda
\vec\pi)\cdot \vec\rho^{\,\nu}
 ~,\displaybreak[0]\\[1ex]
 {\cal L}_{a_1\pi\gamma}  &=  - e\,\frac{1}{m_{a_1}}
F_{\mu\nu} \nonumber \\
 &\quad\mbox{} \times \biggl( 2\bigl[ (\partial^\mu \vec\pi) \times \vec{a}_1^\nu
 -(\partial^\nu \vec\pi) \times \vec{a}_1^\mu\bigr]
 + \vec\pi \times \vec{a}_1^{\mu\nu}  \biggr)
  ~,\displaybreak[0]\\
{\cal L}_{N\Delta\gamma}  &=  ie\frac{G_1}{2m_N} \bar{\Psi}_\Delta^\mu\,
T_z^\dagger \gamma_5\gamma^\nu\, \Psi F_{\mu\nu}
\nonumber\\
  &\quad\mbox{} + e\,\frac{G_2}{4m_N^2}\left( \partial^\nu
\bar{\Psi}_\Delta^\mu\right) T_z^\dagger \gamma_5\Psi F_{\mu\nu} + \ \ \hc~,
\label{LagNDgamg}
\end{align}
\end{subequations}
where $F_{\mu\nu} \equiv \partial_\mu A_\nu - \partial_\nu A_\mu$ with $A_\mu$
denoting the electromagnetic field and $\vec{a}_1^{\mu\nu} \equiv
\partial^\mu \vec{a}_1^\nu - \partial^\nu \vec{a}_1^\mu $. $e$ is the
proton charge; $\hat e = (1 + \tau_z)/2$ and $\hat\kappa = [1.79(1+\tau_z)/2 -
1.93(1-\tau_z)/2]$ are the charge and magnetic moment operators of the nucleon,
respectively. The coupling constants $g_{\omega\pi\gamma}= 0.374$ and
$g_{\rho\pi\gamma} = 0.125$ are fixed from the decay of the $\omega$- and
$\rho$-meson into $\pi^0 + \gamma$, respectively. The signs of these coupling
constants are consistent with those determined from the study of pion
photo-production in the $1$\,GeV region~\cite{Garc}.
$\varepsilon_{\mu\alpha\lambda\nu}$ is the totally antisymmetric Levi-Civita
tensor with $\varepsilon^{0123} = + 1$. The Lagrangian ${\cal
L}_{a_1\pi\gamma}$ is obtained from ${\cal L}_{a_1\pi\rho}$ in Ref.~\cite{Wess}
by combining it with the vector-dominance model.

The propagators required for constructing $M^\mu_s$, $M^\mu_u$, and $M^\mu_t$
are
\begin{subequations}
\begin{align}
 \Delta_\pi(q) & = \frac{1}{ q^2 - m_\pi^2 }
 ~,\\[1ex]
 D_v^{\mu\nu}(q) & =  - \frac{g^{\mu\nu} - q^\mu q^\nu / m_v^2}{
                               q^2 - m_v^2 }~,\quad\text{for}\quad
                               v=\rho,\omega, a_1
 ~,\displaybreak[0]\\[1ex]
 S_N(p)& =  \frac{1}{\fs{p}-m_N}
 ~,\\[1ex]
 S_\Delta^{\mu\nu}(p)&  = \frac{\fs{p} + m_\Delta}{p^2 - m_\Delta^2}
 \nonumber \\
  &  \quad\mbox{}\times \left[- g^{\mu\nu} + \frac{1}{3}\gamma^\mu\gamma^\nu +
      \frac{2}{3}\frac{p^\mu p^\nu}{m_\Delta^2} -
      \frac{p^\mu\gamma^\nu - p^\nu\gamma^\mu}{3m_\Delta} \right]
      ~,
\label{propags}
\end{align}
\end{subequations}
where $\Delta_\pi(q)$ denotes the pion propagator and $D_v^{\mu\nu}(q)$ the
vector ($\rho$, $\omega$) and axial-vector ($a_1$) meson propagators. $S_N(p)$
and $S_\Delta^{\mu\nu}(p)$ are the nucleon and Rarita-Schwinger $\Delta$
propagators, respectively; $m_\Delta = 1232$ MeV denotes the mass of the
$\Delta$. Note that for the $s$-channel $\Delta$ resonance contribution, we
have used the dressed $N\Delta\pi$ vertex and the dressed $\Delta$ propagator
according to Ref.~\cite{Oliver} and consistent with the $\pi N$ FSI used.

The amplitudes $M^\mu_s$, $M^\mu_u$, and $M^\mu_t$ constructed from the
preceding Lagrangians are diagrammatically represented in
Fig.~\ref{diagrams1}(a).

Our model for $M^\mu_s$, $M^\mu_u$ and $M^\mu_t$ is supplemented with hadronic
form factors, except for the $s$-channel $\Delta$ contribution where the
dressed vertex is used. So, the $NN\pi$ vertex in the $s$- and $u$-channels and
the $N\Delta\pi$ vertex in the $u$-channel are multiplied by a form factor
\begin{equation}
F_B(\vec{p}\ ^2) = \frac{\Lambda_B^4}{\Lambda_B^4 + (\vec{p}\ ^2+m_B^2)^2} \ ,
\label{FFDIR}
\end{equation}
where $\vec{p}$ denotes the three-momentum of the off-shell baryon. In the
above equation $B$ stands for either the nucleon or $\Delta$ in the
intermediate state. We take $\Lambda_B=1.2$ GeV for both baryons.

The hadronic vertices in the $t$-channel $M^\mu_t$ amplitude are also
supplemented by form factors of the form
\begin{equation}
F_\alpha(q^2) =  \left( \frac{\Lambda_\alpha^2 - m_\alpha^2} {\Lambda_\alpha^2
- q^2}\right)^{n_\alpha} \ , \label{FF2pi}
\end{equation}
where $\alpha=\pi , \rho , \omega , a_1$. We take $\Lambda_\pi=900$ MeV and
$n_\pi=1$ \cite{Janssen} and $\Lambda_\alpha=1850$ MeV and $n_\alpha=2$ as
$\alpha=\rho , \omega , a_1$ \cite{Machleidt}.


\begin{thebibliography}{99}      


\bibitem{Chew} G.\,F. Chew, M.\,L. Goldberger, F.\,E. Low, and Y. Nambu,
               Phys.\ Rev.\ \textbf{106}, 1345 (1957).



\bibitem{Bernard:1991rt}
V.~Bernard, N.~Kaiser, J.~Gasser and U.-G.~Mei{\ss}ner,
Phys.\ Lett.\ \textbf{B268}, 291 (1991).

\bibitem{Bernard:1992nc}
V.~Bernard, N.~Kaiser and U.-G.~Mei{\ss}ner,
Nucl.\ Phys.\  \textbf{B383}, 442 (1992).

\bibitem{BKLM} V. Bernard, N. Kaiser, T.-S. H. Lee, and U.-G. Mei{\ss}ner,
               Phys.\ Rep.\ \textbf{246}, 315 (1994).

\bibitem{Bernard} V. Bernard, N. Kaiser, and U.-G. Mei{\ss}ner,
                  Phys.\ Lett.\ \textbf{B378}, 337 (1996);
                  Phys.\ Lett.\ \textbf{B383}, 116 (1996).

\bibitem{Feuster} T. Feuster and U. Mosel,
                  Phys.\ Rev.\ C\,\textbf{59}, 460 (1999).

\bibitem{Penner} G. Penner and U. Mosel,
                  Phys.\ Rev.\ C\,\textbf{66}, 055211 (2002); C\,\textbf{66}, 055212 (2002).

\bibitem{Scholten} O. Scholten, A.\,Yu. Korchin, V. Pascalutsa, and D. van Neck,
                   Phys.\ Lett.\ \textbf{B384}, 13 (1996).


\bibitem{DHKT}  D. Drechsel, O. Hanstein, S.\,S. Kamalov, and L. Tiator,
                Nucl.\ Phys.\ \textbf{A645}, 145 (1999).

\bibitem{Hanstein} O. Hanstein, D. Drechsel, and L. Tiator,
                   Nucl.\ Phys.\ \textbf{A632}, 561 (1998).


\bibitem{Nozawa} S. Nozawa, B. Blankleider, and T.-S. H. Lee,
                 Nucl.\ Phys.\ \textbf{A513}, 459 (1990);
                 S. Nozawa, T.-S. H. Lee, and B. Blankleider,
                 Phys.\ Rev.\ C\,\textbf{41}, 213 (1990).

\bibitem{Surya} Y. Surya and F. Gross,
                Phys.\ Rev.\ C\,\textbf{53} 2422 (1996).

\bibitem{Sato} T. Sato and T.-S. H. Lee,
               Phys.\ Rev.\ C\,\textbf{54}, 2660 (1996).

\bibitem{Pascalutsa} V. Pascalutsa and J.\,A. Tjon,
                     Phys.\ Rev.\ C\,\textbf{70}, 035209 (2004).



\bibitem{Gross} F. Gross and D.\,O. Riska,
                Phys.\ Rev.\ C\,\textbf{36}, 1928 (1987).

\bibitem{Ohta} K. Ohta,
               Phys.\ Rev.\ C\,\textbf{40}, 1335 (1989).

\bibitem{Antw} C.\,H.\,M.~van~Antwerpen and I.~Afnan,
               Phys.\ Rev.\ C~\textbf{52}, 554 (1995).

\bibitem{HH1} H. Haberzettl,
              Phys.\ Rev.\ C\,\textbf{56}, 2041 (1997).

\bibitem{HH2} H. Haberzettl, C. Bennhold, T. Mart, and T. Feuster,
              Phys.\ Rev.\ C\,\textbf{58}, R40 (1998).

\bibitem{Kvinikhidze}
A.\,N.~Kvinikhidze and B.~Blankleider, Phys.\ Rev.\ C\,\textbf{60}, 044003
(1999); \textit{ibid.}, 044004 (1999).

\bibitem{DW} R.\,M.~Davidson and R.~Workman,
             Phys.\ Rev.\  C\,\textbf{63}, 025210 (2001).

\bibitem{Borasoy:2005zg}
B.~Borasoy, P.\,C.~Bruns, U.-G.~Mei{\ss}ner and R.~Nissler,
Phys.\ Rev.\ C\,\textbf{72}, 065201 (2005).


\bibitem{Usov} A. Usov and O. Scholten,
 Phys.\ Rev.\ C\,\textbf{72}, 025205 (2005).

\bibitem{Nozawa1} S. Nozawa and T.-S. H Lee,
                  Nucl.\ Phys.\ \textbf{A513}, 511 (1990).

\bibitem{Caia} G.\,L. Caia, V. Pascalutsa, J.\,A. Tjon, and L.\,E. Wright,
               Phys.\ Rev.\ C\,\textbf{70}, 032201(R) (2004);
               G.\,L. Caia, L.\,E. Wright, and V. Pascalutsa,
               Phys.\ Rev.\ C\,\textbf{72}, 035203 (2005).


\bibitem{kazes} E.~Kazes, Nuovo Cimento \textbf{13}, 1226 (1959).


\bibitem{Fred} F. de Jong and K. Nakayama,
              Phys.\ Lett.\ \textbf{B385}, 33 (1996).

\bibitem{HH3} H. Haberzettl,
              Phys.\ Rev.\ C\,\textbf{62}, 034605 (2000).

\bibitem{NL} K. Nakayama and W.\,G. Love,
             Phys.\ Rev.\ C\,\textbf{72}, 034603 (2005).

\bibitem{Oliver} O. Krehl, C. Hanhart, S. Krewald, and J. Speth,
            Phys.\ Rev.\ C\,\textbf{62}, 025207 (2000).

\bibitem{schweber} S.\,S. Schweber, \textit{An Introduction to Relativistic
                    Quantum Field Theory}, (Harper \& Row, New York, NY, 1962;
                    reprinted by Dover, Minealo, NY, 2005).

\bibitem{Mainz1} G. Blanpied \textit{et al.},
                 Phys.\ Rev.\ Lett.\ \textbf{79}, 4337 (1998).

\bibitem{Mainz2} R. Beck \textit{et al.},
                 Phys.\ Rev.\ C\,\textbf{61}, 035204 (2000).


\bibitem{txsc} E. Mazzucato \textit{et al.}, Phys.\ Rev.\ Lett.\ \textbf{57}, 3144 (1986);
               R. Beck \textit{et al.}, Phys.\ Rev.\ Lett.\ \textbf{65}, 1841 (1990);
               M. Fuchs \textit{et al.}, Phys.\ Lett.\ \textbf{B368}, 20 (1996);
               J. C. Bergstrom \textit{et al.}, Phys.\ Rev.\ C\,\textbf{53}, 1052 (1996);
               A. Schmit \textit{et al.}, Phys.\ Rev.\ Lett.\ \textbf{87}, 232501 (2001);
               B.\,B. Govorkov \textit{et al.}, Sov.\ J. Nucl.\ Phys.\ \textbf{6}, 370 (1968);
               W. Hitzeroth \textit{et al.}, Nuovo Cimento A \textbf{60}, 467 (1969);
               J. Ahrens \textit{et al.}, Phys.\ Rev.\ Lett.\ \textbf{84}, 5950 (2000);
               R. G. Vasilkov \textit{et al.}, JETP \textbf{10}, 7 (1960);
               F. H\"arter, PhD Thesis, Johannes-Gutenberg Universit\"at Mainz (1996);
               M. MacCormick \textit{et al.}, Phys.\ Rev.\ C\,\textbf{53}, 41 (1996).


\bibitem{dxsc1} D. Menze, W. Pfeil, and R. Wilcke,
                ZAED Compilation of Pion Photoproduction Data,
                University of Bonn, 1977;
                M. Yoshioka \textit{et al.}, Nucl.\ Phys.\ \textbf{B168}, 222 (1980);
                J.-L. Faure \textit{et al.}, Nucl.\ Phys.\ \textbf{A424}, 383 (1984);
                R. Beck \textit{et al.}, Phys.\ Rev.\ Lett.\ \textbf{78}, 606 (1997);
                J. Ahrens \textit{et al.}, Eur.\ Phys.\ J. \textbf{A21}, 323 (2004);
                A. Shafi \textit{et al.}, Phys.\ Rev.\ C\,\textbf{70}, 035204 (2004);

\bibitem{js72}  H.\,F.~Jones and M.\,D.~Scadron,
                Ann.\ of Phys.\ \textbf{81}, 1 (1972).

\bibitem{Machleidt} R. Machleidt, K. Holinde, and Ch.\ Elster,
                    Phys.\ Rep.\ \textbf{149}, 1 (1987).

\bibitem{Janssen} G. Janssen, K. Holinde, and J. Speth,
                Phys.\ Rev.\ Lett.\ \textbf{73}, 1332 (1994);
                Phys.\ Rev.\ C\,\textbf{49}, 2763 (1994).

\bibitem{Wess} J. Wess and B. Zumino,
               Phys.\ Rev.\ \textbf{163}, 1727 (1967).

\bibitem{Garc} H. Garcilazo and E. Moya de Guerra,
               Nucl.\ Phys.\ \textbf{A562}, 521 (1993).
\end{thebibliography}
\end{document}